\documentclass[a4paper,11pt,superscriptaddress,preprintnumbers,floatfix]{article}
\pdfoutput=1 
\usepackage{jheppub} 
\allowdisplaybreaks
\usepackage{tabularx}
\usepackage{tikz}
\usetikzlibrary{arrows.meta,calc,positioning,decorations.pathreplacing,shapes.geometric}

\newcommand{\bdim}{\chi}   % TEBD bond dimension
\newcommand{\Lcut}{L_{\mathrm{cut}}}
\newcommand{\Cc}{\mathcal{C}}      % charge-conjugation quantum number
\newcommand{\Cop}{\mathcal{C}} % lattice charge-conjugation operator

\begin{document}

\title{Light-cone distributions of vector and scalar bound states in the Schwinger model from tensor networks}

\author[a,b]{Juan Hormaza}
\author[c]{and Wenyang Qian}
\emailAdd{juandavid.hormaza.pantoja@rai.usc.es}
\emailAdd{wqian@ccnu.edu.cn}
\affiliation[a]{Instituto Galego de F\'isica de Altas Enerx\'ias IGFAE, Universidade de Santiago de Compostela, E-15782 Galicia-Spain}
\affiliation[b]{Universidad Nacional de Colombia, Manizales, Colombia.}
\affiliation[c]{Institute of Particle Physics and Key Laboratory of Quark and Lepton Physics (MOE), Central China Normal University, Wuhan, 430079, Hubei, China}

\abstract{
Light-cone distributions encode the partonic structure of hadrons, but they involve real-time correlations that Euclidean lattice methods cannot access directly. We compute the parton distribution functions (PDFs) and light-cone distribution amplitudes (LCDAs) of the vector and scalar meson bound states of the massive Schwinger model with tensor networks. Using the Kogut-Susskind lattice Hamiltonian with the gauge field eliminated through Gauss's law, we prepare the vacuum and the mesons as matrix product states and identify the vector and scalar mesons through their charge conjugation and momentum. The gauge-invariant light-cone correlators that define both distributions are evaluated directly in Minkowski space by real-time evolution. The two mesons exhibit qualitatively different light-cone structure, dictated by charge conjugation. The vector meson has a single-peaked PDF and a symmetric LCDA, while the scalar meson has a double-peaked PDF and an antisymmetric LCDA, whose vanishing zeroth moment requires a normalization through its first moment. The decay constants obtained from local matrix elements and from the moments of the LCDAs agree, and approach predictions from nonrelativistic model as the fermion mass increases. For the relatively large masses considered in this work, the PDFs closely follow the squared LCDAs, showing that the mesons are dominated by their valence component. Our results demonstrate tensor networks as a practical route to light-cone distributions of ground-state as well as excited hadrons in low-dimensional gauge theories.
}
\maketitle

%% ============================================================================
\section{Introduction}
\label{sec:intro}
%% ============================================================================
Understanding the internal structure of hadrons in terms of their quark and gluon constituents is a central goal of nuclear and particle physics, and a major motivation for the future Electron-Ion Collider~\cite{AbdulKhalek:2021gbh}. Much of this structure is encoded in light-cone correlation functions, i.e., hadronic matrix elements of quark and gluon fields separated along a light-like direction, which arise in high-energy processes where the hadron is probed near the light cone~\cite{Collins:1981uw,Collins:2011zzd}. Two prominent examples and useful distributions are the parton distribution functions (PDFs) and light-cone distribution amplitudes (LCDAs). PDFs are forward hadron matrix elements that are interpreted as number densities of partons carrying a fraction $x$ of the hadron's longitudinal momentum. Through collinear factorization~\cite{Collins:2011zzd}, they enter the cross sections of hard inclusive processes, such as deep-inelastic scattering and Drell-Yan production,  and they are extracted from global fits to a wide range of data~\cite{Hou:2019efy,NNPDF:2021njg}. LCDAs, on the other hand, are hadron-to-vacuum matrix elements that encode the probability amplitudes for finding the hadron in its valence Fock sector with the partons carrying momentum fractions $u$ and $1-u$~\cite{Lepage:1980fj,Efremov:1979qk,Chernyak:1983ej}. They govern hard exclusive processes, such as meson form factors at large momentum transfer and exclusive meson decays~\cite{Beneke:1999br}. Both the PDFs and LCDAs are intrinsically nonperturbative. While PDFs are well constrained by data, whereas LCDAs are much less constrained by experiment; a first-principles determination of their full momentum-fraction dependence therefore remains an important goal.
 
Both PDF and LCDA are defined through operators separated along the light cone; therefore, they cannot be computed directly in Euclidean lattice QCD~\cite{RQCD:2019osh, Braun:2006dg, Detmold:2025lyb}. Over the last decade, indirect approaches based on equal-time correlators of boosted hadrons, such as quasi- and pseudo-distributions, have made substantial progress for both PDFs~\cite{Ji:2013dva,Ji:2020ect,Radyushkin:2017cyf,Cichy:2018mum,Constantinou:2020hdm} and LCDAs~\cite{Zhang:2017bzy,Hua:2020gnw,Gao:2022vyh}, by requiring large hadron momenta and a perturbative matching to the light cone. Complementarily, the Hamiltonian formulation offers an alternative route by giving access to real-time evolution and therefore, in principle, enables direct computation of light-cone correlators in Minkowski space without a sign problem~\cite{Barata:2026gee}. This possibility has motivated various proposals for computing parton physics on quantum computers~\cite{Lamm:2019uyc,Mueller:2019qqj,Kreshchuk:2020dla,Kreshchuk:2020aiq,Echevarria:2020wct,Li:2021kcs,Qian:2021jxp,Li:2022lyt} and, for low-dimensional theories, with tensor-network methods~\cite{Orus:2013kga,schollwock2011density,Cirac:2020obd,Banuls:2019bmf,Banuls:2022vxp, Kang:2025xpz}, which provide quasi-exact access to ground and low-lying states and their real-time evolution~\cite{Vidal:2003lvx,Paeckel:2019yjf}.
 
The Schwinger model, or quantum electrodynamics in 1+1 dimensions~\cite{Schwinger:1962tp}, is a natural testing ground for such approaches. It shares many key features with QCD, including confinement, a non-trivial vacuum, chiral-symmetry breaking, and a spectrum of fermion-antifermion bound states~\cite{Coleman:1975pw,Coleman:1976uz}, while remaining analytically solvable in the massless limit and numerically tractable. Its spectrum has been studied with strong-coupling and series expansions~\cite{Banks:1975gq, Hamer:1997dx,Sriganesh:1999ws}, in light-front quantization~\cite{Bergknoff:1976xr,Eller:1986nt,Harada:1995tv}, and with high precision with matrix product states (MPS)~\cite{Byrnes:2002gj,Banuls:2013jaa,Buyens:2013yza,Funcke:2019zna} including the identification of composite-particle states~\cite{Itou:2023img}. Its real-time dynamics, from string breaking to meson scattering, has been simulated with tensor networks~\cite{Pichler:2015yqa,Buyens:2015tea,Papaefstathiou:2024zsu} and on quantum hardware~\cite{Martinez2016, Klco:2018kyo,Kokail:2018eiw,Nguyen:2021hyk,Farrell:2024fit,Farrell:2023fgd,Angelides:2023noe}. Recently, the PDF of its lightest meson, the vector meson, has been computed directly from light-cone correlators with tensor networks~\cite{Banuls:2024oxa,Banuls:2025wiq} as well as quantum hardware~\cite{Chen:2025zeh}; quasi-distributions and generalized parton distributions have also been studied with related techniques~\cite{Grieninger:2024cdl,Li:2026thq}. Analogous approaches have also been applied to purely fermionic models, such as the Nambu--Jona-Lasinio model~\cite{Li:2022lyt,Kang:2025xpz}. However, the LCDAs of the Schwinger model, as well as the partonic structure of its excited mesons, have not yet been studied.
 
In this work, we extend these studies in three directions. First, we compute LCDAs alongside PDFs within the same framework: both follow from the same light-cone bilocal operator, evaluated in forward and in vacuum-to-meson matrix elements, respectively. Second, we study not only the vector meson but also the scalar meson, the lowest excitation that is even under charge conjugation $\Cc$, i.e. that carries the same $\Cc$ quantum number as the vacuum, whereas the vector meson is $\Cc$-odd. In the finite-volume spectrum, the scalar meson appears only as a highly excited state and requires careful state identification. Third, we exploit the relation between the two distributions: for a pure fermion-antifermion state the PDF is the square of the LCDA, so that comparing them probes the Fock-space content of the mesons. Through our study, we find that the two mesons have qualitatively different light-cone structure: the vector meson has an symmetric LCDA under $u \to 1-u$ and a single-peaked PDF, while the scalar meson has an antisymmetric LCDA and a double-peaked PDF, where the node of the LCDA at $u=1/2$ producing the central dip in the PDF. In the heavy-fermion regime studied here, both mesons are dominated by their valence component.
 
The paper is organized as follows. In Sec.~\ref{sec:lc} we define both light-cone distributions and discuss their properties. In Sec.~\ref{sec:tn}, we describe the lattice Hamiltonian and the tensor-network methods used to prepare the meson states and to evaluate the light-cone correlators. In Sec.~\ref{sec:results}, we present numerical results are presented using \texttt{iTensor} library~\cite{itensor}, and we summarize the paper in Sec.~\ref{sec:conclusion}.

%% ============================================================================
\section{Light-cone distributions of mesons}
\label{sec:lc}
%% ============================================================================
PDFs and LCDAs are the standard light-cone correlation functions of a hadron. They are not observables in the strict sense, but universal, process-independent matrix elements that enter observables through factorization theorems~\cite{Collins:2011zzd,Lepage:1980fj}. In $3+1$ dimensions, with a light-like vector $n^\mu$ and a hadron of momentum $P$, the unpolarized quark PDF is defined by~\cite{Collins:2011zzd}
\begin{align}
    f_h(\xi)=\int\frac{d\lambda}{4\pi}\,e^{-i\xi\lambda\,n\cdot P}\,\langle h(P)|\bar\psi(\lambda n)\,W(\lambda n\leftarrow0)\,(n\cdot\gamma)\,\psi(0)|h(P)\rangle\,,
    \label{eq:pdf_4d}
\end{align}
where $W$ is the Wilson line along $n$, the hadron spin is averaged, and the vacuum contribution is subtracted. For $\xi>0$, $f_h(\xi)$ is the number density of quarks carrying momentum fraction $\xi$, while for $\xi<0$, $-f_h(\xi)$ gives the antiquark density at momentum fraction $|\xi|$.

The leading-twist LCDA of a meson $h$ is defined by the vacuum-to-meson matrix element of the same bilocal operator, with a Dirac structure $\Gamma_h$ that depends on the meson~\cite{Lepage:1980fj,Chernyak:1983ej,Braun:2003rp,Ball:1996tb,Ball:1998sk,Cheng:2005nb,Yang:2007zt},
\begin{align}
    i \mathsf{f}_h\,\phi_h(u)=\int\frac{d\lambda}{2\pi}\,e^{iu\lambda\,n\cdot P}\,\langle \Omega|\bar\psi(\lambda n)\,W(\lambda n\leftarrow0)\,\Gamma_h\,\psi(0)|h(P)\rangle\,,
    \label{eq:lcda_4d}
\end{align}
where $|\Omega \rangle$ is the vacuum state, $u$ is the momentum fraction of the valence parton, and $\mathsf{f}_h$ is the decay constant. 
Three Dirac structures are allowed at leading twist: $n\cdot\gamma$, $(n\cdot\gamma)\gamma_5$, and $i\sigma^{n\nu_\perp}$. The first two project the quark-antiquark pair onto total helicity zero and the third onto helicity $\pm1$. 

Charge conjugation further constrains the LCDAs of neutral mesons. If $\eta_\Gamma$ is the charge-conjugation parity of the local operator $\bar\psi\Gamma\psi$ ($-1$ for $\gamma^\mu$ and $\sigma^{\mu\nu}$, $+1$ for $\gamma^\mu\gamma_5$) and $\eta_{\Cc}$ that of the meson, then $\phi_h(1-u)=\eta_\Gamma\eta_{\Cc}\,\phi_h(u)$. There can be two cases. For $\eta_\Gamma\eta_{\Cc}=+1$, the LCDA is symmetric about $u=1/2$, the local matrix element is allowed, and the LCDA is normalized as $\int_0^1du\,\phi_h(u)=1$, so that $\mathsf{f}_h$ is the decay constant, $\langle\Omega|\bar\psi\,\Gamma_h\psi|h(P)\rangle=i\mathsf{f}_h\,n\cdot P$ (up to polarization factors for vector mesons). This covers the familiar mesons: the $\pi^0$ and $\eta_c$ ($\eta_{\Cc}=+1$), and the $\rho^0$ and $J/\psi$ ($\eta_{\Cc}=-1$)~\cite{Braun:2003rp,Ball:1996tb,Ball:1998sk}. For $\eta_\Gamma\eta_{\Cc}=-1$, the LCDA is antisymmetric, such that its zeroth moment and hence the local matrix element vanishes, and we normalize by its first moment, $\int_0^1du\,(2u-1)\,\phi_h(u)=1$, so that $\mathsf{f}_h$ is defined by the matrix element of the twist-two operator with one covariant derivative. Examples are the neutral scalar mesons ($\eta_{\Cc}=+1$)~\cite{Cheng:2005nb} and the $P$-wave scalar quarkonia~\cite{Braguta:2008qe,Li:2026fgc}, as well as the $b_1$ meson ($\eta_{\Cc}=-1$)~\cite{Yang:2007zt}.

In $1+1$ dimensions, the three leading-twist structures reduce to one. With one spatial dimension, there are no rotations, hence no spin, no helicity, and no transverse direction, and both distributions depend only on the longitudinal momentum fraction. The Dirac algebra is spanned by $\{1,\gamma^0,\gamma^1,\gamma_5\}$ with $\gamma_5=\gamma^0\gamma^1$, so that the tensor reduces to $\sigma^{\mu\nu}=(i/2)[\gamma^\mu, \gamma^\nu ]=i\epsilon^{\mu\nu}\gamma_5$ and the axial structure to the vector one, $\gamma^\mu\gamma_5=-\epsilon^{\mu\nu}\gamma_\nu$ with $\epsilon^{01}=+1$. Hence, $(n\cdot\gamma)\gamma_5=-\,n\cdot\gamma$, and no transverse structure exists. A single leading-twist structure, $n\cdot\gamma$, remains, and therefore defines one PDF and one LCDA for every meson. Since $\bar\psi\gamma^\mu\psi$ is $\Cc$-odd, the rule above becomes $\phi(1-u)=-\eta_{\Cc}\,\phi(u)$. In the absence of a background electric field, charge conjugation and parity are exact symmetries of the model, and for the mesons considered here parity coincides with $\Cc$~\cite{Coleman:1976uz}, so $\eta_{\Cc}$ is the only independent label for mesons. The $\Cc$-odd meson therefore has a symmetric LCDA and can couple to the local current, and the $\Cc$-even meson an antisymmetric LCDA and a vanishing local matrix element. Throughout the rest of this paper, we work in $1+1$ dimensions. 

\subsection{Light-cone bilocal operator}
\label{sec:bilocal}
We use light-cone coordinates $v^\pm=(v^0\pm v^1)/\sqrt{2}$ and the light-like vector $n^\mu=(1,1)/\sqrt{2}$, for which $n\cdot v=v^-$ and $n\cdot\gamma=\gamma^-=(\gamma^0-\gamma^1)/\sqrt{2}$. All distributions considered here follow from a single gauge-invariant bilocal operator
\begin{align}
    \mathcal{O}(\lambda)=\bar{\psi}(\lambda n)\,W(\lambda n\leftarrow 0)\,\gamma^-\psi(0)\,,
\qquad
W(\lambda n\leftarrow 0)=\exp\left[-ig\int_0^\lambda d\tau\,n\cdot A(\tau n)\right]\,.
    \label{eq:bilocal}
\end{align}
where the fermion fields are separated along the light cone and connected by a Wilson line. We consider mesons $h=V,S$ at rest, $P^\mu=(M_h,0)$, so that $n\cdot P=M_h/\sqrt{2}$. Taking $\lambda n=(z,z)$, then $\lambda\,n\cdot P=M_hz$. The PDF and the LCDA are then Fourier transforms of the forward and the vacuum-to-meson matrix elements of $\mathcal{O}$, respectively.

\subsection{Parton distribution function}
\label{sec:pdf}

The PDF of the fermion in meson $h$ is~\cite{Collins:2011zzd}
\begin{align}
    f_h(\xi)=\int\frac{d\lambda}{4\pi}\,e^{-i\xi\lambda\,n\cdot P}\,
    \Big[\langle h(P)|\mathcal{O}(\lambda)|h(P)\rangle-\langle\Omega|\mathcal{O}(\lambda)|\Omega\rangle\Big]\,,
    \label{eq:pdf}
\end{align}
where the subtraction removes the disconnected contribution from the vacuum $|\Omega\rangle$. For $0\le\xi\le1$, $f_h(\xi)$ is the number density of fermions carrying the light-cone momentum fraction $\xi$, summed over all Fock components of the meson. For $-1\le\xi<0$, $-f_h(\xi)$ is the density of antifermions with momentum fraction $|\xi|$. For a charge-conjugation eigenstate, the fermion and antifermion densities coincide, so that $f_h(-\xi)=-f_h(\xi)$ and the net fermion number vanishes, $\int_{-1}^{1}d\xi\,f_h(\xi)=0$.

The zeroth moment of the PDF counts the fermions in the meson, and defined as
\begin{align}
n_h=\int_0^1d\xi\,f_h(\xi)=P_2+2P_4+3P_6+\dots\ge1\,,
\label{eq:nh}
\end{align}
where $P_{2k}$ is the probability of the $k$-fermion-antifermion pairs, $\sum_k P_{2k}=1$. It equals one for a pure fermion-antifermion state and exceeds one if higher Fock components are present. In $1+1$ dimensions, the gauge field carries no light-cone momentum, since its contribution vanishes identically~\cite{Einhorn:1976uz,Jia:2018qee}. The total momentum of the hadron is therefore carried entirely by fermions and antifermions, $\int_{-1}^{1}d\xi\,\xi\,f_h(\xi)=1$, so that for an odd PDF, its first moment is
\begin{align}
    \langle \xi \rangle_h = \int_0^1d\xi\,\xi\,f_h(\xi)=\frac12\,,
    \label{eq:momentum_sum_rule}
\end{align}
independent of the Fock-space content of the meson, which is an exact constraint to verify the numerical results.

%% ============================================================================
\subsection{Light-cone distribution amplitude}
\label{sec:lcda}
%% ============================================================================

The LCDA follows from the vacuum-to-meson matrix element of the bilocal operator,
\begin{align}
    R_h(u)\equiv\int\frac{d\lambda}{2\pi}\,e^{iu\lambda\,n\cdot P}\,
    \langle\Omega|\mathcal{O}(\lambda)|h(P)\rangle\equiv i\mathsf{f}_h\,\phi_h(u)\,,
    \label{eq:lcda}
\end{align}
where $u$ and $1-u$ are the momentum fractions of the two valence constituents. In light-front quantization, the vacuum contains no particles and $\mathcal{O}$ annihilates at most one fermion-antifermion pair, so $R_h$ probes only the two-particle component of the meson and is proportional to its light-cone wave function. Here, no vacuum subtraction is needed, as the matrix element connects orthogonal states. In $1+1$ dimensions, there is no transverse momentum to integrate over, therefore, $R_h(u)$ is also directly proportional to the two-particle light-front wave function. The constant $\mathsf{f}_h$ is a local vacuum-to-meson matrix element and thus a physical quantity, while $\phi_h$ carries the shape; the normalization of $\phi_h$ is chosen such that $\mathsf{f}_h$ is the lowest non-vanishing moment of $R_h$.

Taking the inverse of Eq.~\eqref{eq:lcda} gives $\langle\Omega|\mathcal{O}(\lambda)|h(P)\rangle=n\cdot P\int_0^1du\,e^{-iu\lambda n\cdot P}R_h(u)$, whose expansion in $\lambda$ relates the zeroth moment to the local current and the first moment to the local operator with one covariant derivative~\cite{Chernyak:1983ej,Braun:2006dg}. The vector meson ($\eta_{\Cc}=-1$) has a symmetric LCDA, normalized as $\int_0^1du\,\phi_V(u)=1$. The scalar meson ($\eta_{\Cc}=+1$) has an antisymmetric LCDA, whose zeroth moment vanishes because the $\Cc$-odd current cannot connect the $\Cc$-even vacuum to a $\Cc$-even meson, so we normalize it through its first moment, $\int_0^1du\,(2u-1)\,\phi_S(u)=1$~\cite{Cheng:2005nb,Yang:2007zt,Braguta:2008qe}. Consequently, the decay constants are obtained from either the normalization or the local matrix elements at $\lambda=0$,
\begin{align}
    \mathsf{f}_V&=\Big|\int_0^1du\,R_V(u)\Big|=\frac{1}{n\cdot P}\,\big|\langle\Omega|\mathcal{O}(0)|V(P)\rangle\big|\,,
    \label{eq:fV}\\
    \mathsf{f}_S&=\Big|\int_0^1du\,(2u-1)\,R_S(u)\Big|=\frac{2}{(n\cdot
P)^2}\,\Big|\frac{d}{d\lambda}\Big[e^{i\lambda n\cdot
P/2}\langle\Omega|\mathcal{O}(\lambda)|S(P)\rangle\Big]_{\lambda=0}\Big|\,.
    \label{eq:scalar_norm}
\end{align}
The phase $e^{i\lambda n\cdot P/2}$ centers the moment at $u=1/2$ and can be dropped for an exact $\Cc$-even state, for which $\int_0^1du\,R_S=0$. Since the overall phase of $|h(P)\rangle$ is arbitrary, we take absolute values for $\mathsf{f}_h$.

%% ============================================================================
\section{Tensor-network simulation of the lattice Schwinger model}
\label{sec:tn}
%% ============================================================================
\subsection{The Schwinger model}
\label{sec:hamiltonian}
%% ============================================================================

The Schwinger model is quantum electrodynamics in 1+1 dimensions, whose Lagrangian density is defined as,
\begin{align}
\mathcal{L}=\bar\psi\left(i\gamma^\mu D_\mu-m\right)\psi-\frac14F_{\mu\nu}F^{\mu\nu}\,,
\qquad D_\mu=\partial_\mu+igA_\mu\,,
\end{align}
with gauge coupling $g$ and fermion mass $m$. In the continuum, the Schwinger model in the temporal gauge $(A_0=0)$ is described by the Hamiltonian
\begin{align}
    H_{\mathrm{cont}} = \int dx \,\left[ -i\,\bar{\psi}(x)\gamma^1\big(\partial_x - igA^1(x)\big)\psi(x)
+ m\bar{\psi}(x)\psi(x) + \frac{1}{2}E(x)^2 \right]\,,
\label{eq:H_cont}
\end{align}
where $A_1(x)$ is the spatial gauge potential, $E(x)$ is the conjugate electric field, and $\bar{\psi}(x)=\psi^\dagger(x)\gamma^0$. We choose $\gamma^0=\sigma^z$ and $\gamma^1=i\sigma^y$, which satisfy $\{\gamma^\mu,\gamma^\nu\}=2\eta^{\mu\nu}$, with $\eta^{\mu\nu}=\operatorname{diag}(1,-1)$, so that $\gamma_5 = \gamma^0 \gamma^1=\sigma^x$. Gauge invariance of the model is implemented through Gauss’s law, which imposes a constraint on the Hilbert space, $G(x) = \partial_x E(x) - g\,\psi^\dagger(x)\psi(x)$, and physical states must satisfy $G(x)\,|\Psi_{\mathrm{phys}}\rangle=0$ to ensure that the dynamics remains restricted to the gauge-invariant subspace. 

Importantly, the physics depends only on $m/g$. For $m=0$, the massless Schwinger model is equivalent to a free bosonic theory whose single excitation is the vector meson~\cite{Schwinger:1962tp,Lowenstein:1971fc,Coleman:1975pw,Coleman:1976uz}. In light-front quantization, it becomes a pure fermion-antifermion state with a constant wave function with $M_V=g/\sqrt{\pi}$ and $f_V(\xi)=\phi_V(u)=1$~\cite{Bergknoff:1976xr,Eller:1986nt}. There is no scalar bound state as the $\Cc$-even channel starts at the two-meson threshold. Perturbation theory for small mass expansion is also available~\cite{Adam:1997wt} and validated numerically~\cite{Banuls:2013jaa}.

For $m>0$, the model is interacting and its spectrum consists of fermion-antifermion bound states, the two lowest of which are stable for all $m/g$~\cite{Coleman:1976uz,Adam:1997wt}: the $\Cc$-odd vector meson $V$ and the $\Cc$-even scalar meson $S$, with $M_V<M_S<2M_V$. 
In particular, For $m\gg g$, the fermion and antifermion are nonrelativistic and pair creation is suppressed, so that the mesons are well described by a single pair bound by the electric field between
them. Here, the charges at separation $r$ are connected by a constant field $E=g$ with energy $\frac12E^2|r|$, which gives the linear potential $V(r)=\sigma|r|$ with string tension $\sigma=g^2/2$. In this non-relativistic model (NR model), the meson wave functions are the Airy functions, which offers a good approximation to PDFs, LCDAs, and decay constants when $m \gg g$. See App.~\ref{app:hf_model} for detailed derivations.

%% ============================================================================
\subsection{The lattice Hamiltonian}
%% ============================================================================

Moving from Eq.~\eqref{eq:H_cont}, we discretize the spatial direction using the Kogut-Susskind staggered-fermion  formulation~\cite{Kogut:1974ag,Susskind:1976jm,Banks:1975gq} on a lattice with $N$ sites and spacing $a$. Each site $n$ carries a single-component fermion field $\phi_n$, with $\{\phi_n,\phi^\dagger_{n'}\}=\delta_{nn'}$, and the two components of the Dirac field are distributed over neighboring sites, $\psi(x_r)=(\phi_{2r-1},\phi_{2r})^T/\sqrt{2a}$, so that a unit cell of length $2a$ contains one Dirac fermion.
% The fermionic annihilation and creation operators at site $n$ are denoted by $\phi_n$ and $\phi_n^\dagger$, respectively. 
The resulting lattice Hamiltonian is
\begin{align}
    H_{\mathrm{lat}} = -\frac{i}{2a}\sum_{n=1}^{N-1}\left(\phi_n^\dagger e^{i\theta_n}\phi_{n+1}-\phi_{n+1}^\dagger e^{-i\theta_n}\phi_n\right)+m\sum_{n=1}^{N}(-1)^{n+1}\phi_n^\dagger\phi_n+\frac{ag^2}{2}\sum_{n=1}^{N-1}L_n^2\,.
\label{eq:schwinger_lattice_hamiltonian}
\end{align}
Here, $\theta_n=-agA^1$ is the gauge field on the link connecting sites $n$ and $n+1$, and $L_n$ is the corresponding dimensionless electric-field operator, with $E_n=gL_n$ and $[\theta_n,L_m]=i\delta_{nm}$. Odd (even) sites host fermions (antifermions): an occupied odd site represents a fermion and an empty even site an antifermion, with staggered charge $Q_n=\phi_n^\dagger\phi_n-[1-(-1)^{n+1}]/2$. Gauss's law takes the form $G_n=L_n-L_{n-1}-Q_n $, where $Q_n$ is the staggered charge operator at site $n$, and the physical states satisfy $G_n|\Psi_{\mathrm{phys}}\rangle=0$. For the mass, we take
$m\rightarrow m_{\rm lat}=m-ag^2/8$ to remove the leading lattice artifact of the staggered
discretization~\cite{Dempsey:2022nys,Angelides:2023noe}.

With open boundary conditions, all the gauge fields, i.e., link phases, can be removed by a gauge transformation, and Gauss's law fixes the electric fields in terms of the charges. After applying the Jordan-Wigner transforms for fermions fields,
\begin{align}
    \phi_n=\prod_{k<n}(-i\sigma^z_k)\,\sigma^-_n\,,\qquad \phi_n^\dagger=\prod_{k<n}(i\sigma^z_k)\,\sigma^+_n\,,
    \label{eq:JW}
\end{align}
with Pauli operators $\sigma^x_n,\sigma^y_n,\sigma^z_n$ and $\sigma^\pm_n=(\sigma^x_n\pm i\sigma^y_n)/2$, the lattice Hamiltonian becomes a spin Hamiltonian with long-range interactions~\cite{Banks:1975gq,Banuls:2013jaa},
\begin{align}
    H=\frac{1}{2a}\sum_{n=1}^{N-1}&\left(\sigma^+_n\sigma^-_{n+1}+\mathrm{h.c.}\right)
    +\frac{m}{2}\sum_{n=1}^{N}(-1)^{n+1}\sigma^z_n
    +\frac{ag^2}{2}\sum_{n=1}^{N-1}L_n^2\,,\label{eq:H}\\
    &L_n=L_0+\sum_{k=1}^{n}\left(Q_k+q_k\right)\,,\qquad Q_k=\frac{\sigma^z_k+(-1)^{k+1}}{2} \,.
    \label{eq:Ln}
\end{align}
We set the boundary field $L_0=0$ and work in the zero-charge sector, $\sum_n \sigma^z_n=0$. The static external charges $q_k$ vanish for the Hamiltonian that is used to calculate the bound states; non-vanishing $q_k$ will be used to implement the moving static charges when calculating the Wilson line~\cite{Banuls:2025wiq,Banuls:2024oxa}. The lattice spacing is often quoted through $x=1/(ag)^2$ and the continuum limit corresponds to $x\to\infty$ at fixed $m/g$ and fixed physical volume $\mathcal{V} =Nag=N/\sqrt{x}$.

%% ============================================================================
\subsection{Meson bound states}
\label{sec:dmrg}
\begin{figure}[t]
\centering
\begin{tikzpicture}[
    mps/.style={circle,draw,fill=blue!25,minimum size=4mm,inner sep=0pt},
    mpsj/.style={circle,draw,fill=orange!45,minimum size=4mm,inner sep=0pt},
    mpo/.style={rectangle,draw,fill=red!25,minimum size=4mm,inner sep=0pt},
    x=5.5mm,y=7mm]
  % ---- panel labels (aligned: (a),(b) share x; (b),(c) share y)
  \node[font=\small] at (-1.9,5.0) {(a)};
  \node[font=\small] at (-1.9,2.6) {(b)};
  \node[font=\small] at (16.6,2.6) {(c)};
  % ================= (a) MPS =================
  \node[left] at (1.3,4.6) {$|\Psi\rangle=$};
  \foreach \i/\x/\lab in {1/1.6/1,2/3.2/2,3/4.8/3,5/8.0/N{-}1,6/9.6/N}{
    \node[mps] (a\i) at (\x,4.6) {};
    \draw (a\i)--++(0,-0.6) node[below,font=\scriptsize,inner sep=1pt] {$s_{\lab}$};
    \node[above,font=\scriptsize,inner sep=2pt] at (a\i.north) {$A_{\lab}$};
  }
  \draw (a1)--(a2)--(a3); \draw (a5)--(a6);
  \draw (a3)--(5.9,4.6); \draw (6.9,4.6)--(a5);
  \node at (6.4,4.6) {$\cdots$};
  \node[below,font=\scriptsize,inner sep=1pt] at (2.4,4.6) {$D_1$};
  \node[below,font=\scriptsize,inner sep=1pt] at (4.0,4.6) {$D_2$};
  \node[right,font=\small] at (10.8,4.6) {$s_n\in\{\uparrow,\downarrow\}$,\quad $D_n\le D$};
  % ================= (b) H_k =================
  \foreach \i in {1,...,5}{
    \node[mps] (t\i) at (\i,2) {};
    \node[mpo] (h\i) at (\i,1) {};
    \node[mps] (b\i) at (\i,0) {};
    \draw (t\i)--(h\i)--(b\i);
  }
  \foreach \i in {1,...,4}{
    \pgfmathtruncatemacro{\j}{\i+1}
    \draw (t\i)--(t\j); \draw[thick] (h\i)--(h\j); \draw (b\i)--(b\j);
  }
  \node[left] at (0.5,2) {$\langle\Psi|$};
  \node[left] at (0.5,1) {$H$};
  \node[left] at (0.5,0) {$|\Psi\rangle$};
  \node at (6.9,1) {$+\;w\displaystyle\sum_{j<k}$};
  \draw[thick] (8.3,-0.2)--(8.3,2.2);
  \foreach \i in {1,...,5}{
    \node[mps]  (p\i) at (\i+8.0,1.6) {};
    \node[mpsj] (q\i) at (\i+8.0,0.4) {};
    \draw (p\i)--(q\i);
  }
  \foreach \i in {1,...,4}{
    \pgfmathtruncatemacro{\j}{\i+1}
    \draw (p\i)--(p\j); \draw (q\i)--(q\j);
  }
  \draw[thick] (13.7,-0.2)--(13.7,2.2);
  \node[font=\small] at (14.15,2.1) {$2$};
  \draw[decorate,decoration={brace,amplitude=4pt,mirror}] (0.6,-0.5)--(14.2,-0.5)
     node[midway,below=5pt] {$\langle\Psi|H_k|\Psi\rangle$, minimized over MPS with bond dimension $\leq D$};
  % ================= (c) level ladder =================
  \begin{scope}[shift={(17.6,-0.6)},x=1cm,y=0.85cm]
    \draw[->] (-0.1,0)--(-0.1,3.2) node[above] {\small $E$};
    \draw[thick] (0,0)--(0.8,0) node[right] {\small $\Omega$, $k=0$};
    \draw[thick,blue] (0,1.2)--(0.8,1.2) node[right] {\small $V$, $k=1$};
    \foreach \y in {1.45,1.75,2.1,2.5}{\draw[blue!60] (0,\y)--(0.8,\y);}
    \node[right,blue!70] at (0.8,1.95) {\small $V$, $\langle P^2\rangle>0$};
    \draw[thick,orange!80!black] (0,2.85)--(0.8,2.85) node[right] {\small $S$, $k=n_S$};
  \end{scope}
\end{tikzpicture}
\caption{
Schematic diagrams for sequential computation of excited states. Panel (a) matrix product state for Eq.~\eqref{eq:mps}; (b) functional minimized for the $k$-th excited state as in Eq.~\eqref{eq:penalty}; (c) schematic ordering of the vacuum, vector, and scalar states in the finite open chain.
}
\label{fig:dmrg}
\end{figure}
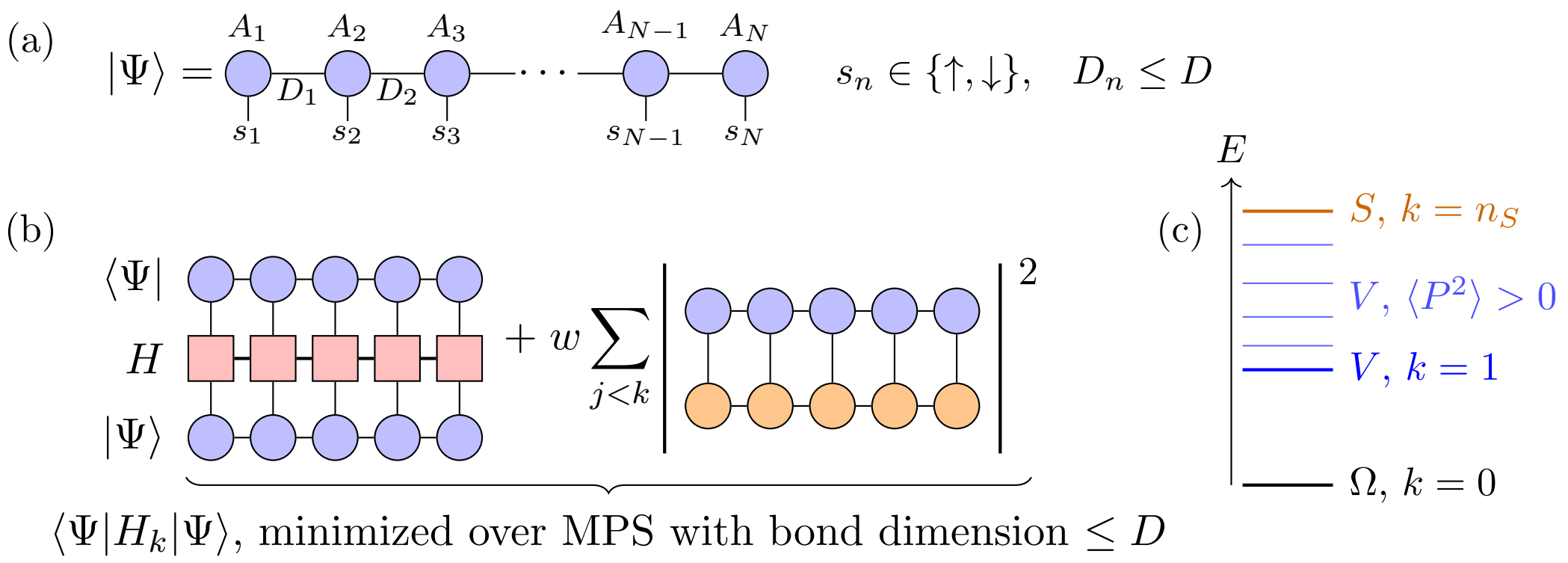

%% ============================================================================
In the Schwinger model, mesons appear as excited eigenstates of the lattice Hamiltonian in the zero-charge sector, and are identified by their quantum numbers. All states are represented as matrix product states (MPS),
\begin{align}
|\Psi\rangle=\sum_{s_1,\dots,s_N}A_1^{s_1}A_2^{s_2}\cdots A_N^{s_N}\,|s_1s_2\cdots s_N\rangle\,,
    \label{eq:mps}
\end{align}
where $s_n\in\{\uparrow,\downarrow\}$ labels the eigenstates of $\sigma^z_n$, i.e. the occupation of site $n$, and $A_n^{s_n}$ is a $D_{n-1}\times D_n$ matrix with $D_0=D_N=1$. The bond dimension $D=\max_nD_n$ controls the entanglement that the state can carry and hence its accuracy, as in Fig.~\ref{fig:dmrg}(a). Since $H$ conserves the total charge, the tensors are block-sparse with respect to the U(1) charge, which restricts all states to the zero-charge sector, $\sum_n\sigma^z_n=0$.

We first obtain the vacuum $|\Psi_0\rangle=|\Omega\rangle$ using the density matrix renormalization group (DMRG) algorithm~\cite{White:1992zz,White:1993zza,schollwock2011density,Schollwock2005} in its MPS formulation, which minimizes $\langle\Psi|H|\Psi\rangle$ by optimizing the tensors $A_n$ site by site. A strong-coupling product state with alternating occupations is used as the initial ansatz for the vacuum. Excited states are then computed sequentially by projecting out the states already found. The $k$-th excited state is obtained as the ground state of
\begin{align}
    H_k = H + w\sum_{j<k}|\Psi_j\rangle\langle\Psi_j| \,,
    \label{eq:penalty}
\end{align}
where $|\Psi_0\rangle=|\Omega\rangle$ is the ground state, $|\Psi_1\rangle,|\Psi_2\rangle,\dots,|\Psi_{k-1}\rangle$ are lower excited states, and the penalty weight $w$ is chosen larger than the energy gap of interest, so that the previously found states are pushed above $|\Psi_k\rangle$ in the spectrum. Our procedure is illustrated in Fig.~\ref{fig:dmrg}(b). The projector terms do not require an MPO (Matrix Product Operator) representation: their contributions to the local effective Hamiltonian are built from the overlaps $\langle\Psi_j|\Psi\rangle$, at a cost that grows linearly with the number of previously computed states. 
We used a fixed penalty weight $w=50\,g$ larger than all gaps of interest and a maximum bond dimension that is gradually increased from $D=10$ to $D=400$, with the truncation threshold lowered from $10^{-6}$ to $10^{-12}$ along the way. To avoid convergence to local minima, a subspace-expansion noise term of strength $10^{-3}$, $10^{-5}$, and $10^{-7}$ is applied in successive blocks of 1000 sweeps and switched off for the final 3000 sweeps.

For each state in the DMRG spectrum, we classify it using two operators in the spectrum: the charge conjugation operator and the discretized momentum operator, similar to Ref.~\cite{Banuls:2013jaa}. With open boundary condition (OBC), neither momentum nor charge conjugation is an exact quantum number but an approximation. The low-lying zero-charge spectrum contains, besides the vector meson at rest, a tower of vector-meson states with nonzero quasi-momentum, i.e. standing waves in the finite volume, which lie below the scalar meson; see Fig.~\ref{fig:dmrg}(c). 
The lattice charge-conjugation operator is define as ~\cite{Banks:1975gq,Banuls:2013jaa}
\begin{align}
    \Cop = T\prod_{n=1}^{N}\sigma^x_n \,,
    \label{eq:charge_conj}
\end{align}
where $T$ is the cyclic translation by one site. The translation exchanges the fermion and antifermion sublattices and the global spin flip performs the particle-hole exchange, so that $Q_n\to-Q_{n+1}$ and $\phi_n\to\phi^\dagger_{n+1}$ up to phases. With OBC, $\Cop$ does not commute with $H$, but it provides an approximation: $\langle\Cop\rangle\simeq+1$ for $\Cc$-even states (the vacuum and the scalar branch) and $\langle\Cop\rangle\simeq-1$ for $\Cc$-odd states (the vector branch). 

Next is the momentum operator. Discretizing $\hat P=\int dx\,\psi^\dagger(x)\,i\partial_x\psi(x)$ and applying Jordan-Wigner transform in Eq.~\eqref{eq:JW} gives
\begin{align}
    P = -\frac{i}{2a}\sum_{n=1}^{N-2}\left[\sigma_n^-\sigma^z_{n+1}\sigma_{n+2}^+-\mathrm{h.c.}\right]\,.
    \label{eq:OP}
\end{align}
We use the expectation value $\langle P^2\rangle$, computed with an exact MPO for $P^2$, to assign a pseudo-momentum to each state, which grows along the tower of vector mesons. In this study, we identify the vector meson as the lowest excitation with $\langle\Cop\rangle\simeq-1$, and the scalar meson as the lowest excitation above the vacuum with $\langle\Cop\rangle\simeq+1$. Both states have $\langle P^2\rangle$ close to the minimal value of the spectrum, i.e.\ they are the mesons nearest to rest.

%% ============================================================================
\subsection{Light-cone real-time correlators}
\label{sec:lightcone}
%% ============================================================================

A Dirac field is reconstructed from two neighboring sites, $\psi_r=(\phi_{2r-1},\phi_{2r})^T$. In our representation of the Dirac matrices, the spinor structure of Eq.~\eqref{eq:bilocal} is $\gamma^0\gamma^-=(I-\sigma^x)/\sqrt{2}$. Placing the initial cell at sites $n_0$ and $n_0+1$ in the middle of the chain, and the final cell at sites $n_0+d$ and $n_0+d+1$ with even values of $d$, the bilocal operator decomposes into four staggered contributions,
\begin{align}
    \mathcal{O}(d)=\mathcal{O}_{0,0}(d)-\mathcal{O}_{0,1}(d)-\mathcal{O}_{1,0}(d)+\mathcal{O}_{1,1}(d)\,,
    \qquad
    \mathcal{O}_{\alpha,\beta}(d)=\phi^\dagger_{n_0+d+\alpha}(t_d)\,\mathcal{W}_d\,\phi_{n_0+\beta}\,,
    \label{eq:four_terms}
\end{align}
where $\alpha,\beta\in\{0,1\}$ label the components of the final and initial spinor, and $\mathcal{W}_d$ denotes the lattice Wilson line along the discretized light-cone path. Owing to $\psi_\alpha=\phi/\sqrt{2a}$ on each sublattice and $\gamma^0\gamma^-=(I-\sigma^x)/\sqrt2$, the discretized bilocal operator $\mathcal{O}(d)$ equals $2\sqrt2\,a$ times the continuum operator of Eq.~\eqref{eq:bilocal} at $\lambda n=(z_d,z_d)$.
Then, we compute light-cone correlators
\begin{align}
    \mathcal{M}_h(d)=\langle h|\mathcal{O}(d)|h\rangle-\langle\Omega|\mathcal{O}(d)|\Omega\rangle\,,
    \qquad
    \mathcal{A}_h(d)=\langle\Omega|\mathcal{O}(d)|h\rangle\,,
    \label{eq:MA}
\end{align}
for the PDF and the LCDA, respectively. The two involve the same operator and differ only in the bra state. 

The Wilson line $\mathcal{W}_d$ along the light cone cannot be applied directly on a spatial lattice. Following the technique first introduced in Ref.~\cite{Banuls:2024oxa}, we replace the light-cone path by a staircase of alternating real-time evolutions and spatial transports. Each step advances the path by one Dirac cell, $\delta z=2a$, and evolves in time by $\Delta t=2a$. After $|d|/2$ steps the path reaches the light-cone point $(z_d,z_d)$, with $z_d=t_d=d a$. Since the gauge field has been integrated out, the spatial parallel transporters cannot be applied as operators. Instead, we use a static charge $q_k$ in Gauss's law, Eq.~\eqref{eq:H}, which compensates the charge removed by $\phi_{n_0+\beta}$ and is moved along the path: shifting it by one Dirac cell changes the flux on the traversed links by one unit, as the action of the link operators $e^{\pm i\theta_n}$ would, and the temporal transporter becomes the time evolution with the Hamiltonian $H_j$ that contains the static charge at its current position. For $d>0$,
\begin{align}
    \mathcal{O}_{\alpha,\beta}(d)\;\longrightarrow\;
    e^{iHt_d}\;\sigma^+_{n_0+d+\alpha}\prod_{k<n_0+d+\alpha}(i\sigma^z_k)\;
    e^{-iH_{d/2}\Delta t}\cdots e^{-iH_{1}\Delta t}\;
    \prod_{k<n_0+\beta}(-i\sigma^z_k)\,\sigma^-_{n_0+\beta}\,,
    \label{eq:zigzag}
\end{align}
where during the $j$-th step the static charge sits at site $n_0+2j-1$. For $d<0$ the path is constructed in the opposite direction with $\Delta t\to-\Delta t$. 
Note only the ket is evolved in time. Since $\langle h|$ and $\langle\Omega|$ are eigenstates of $H$, the evolution $e^{iHt_d}$ acting on the bra reduces to the phase $e^{iE_ht_d}$ in $\langle h|\cdots|h\rangle$ and $e^{iE_0t_d}$ in $\mathcal{A}_h$, with the DMRG energies, and each correlator is obtained as the overlap of the evolved MPS with the bra MPS. The four components $\mathcal{O}_{\alpha,\beta}$ are computed separately and combined according to Eq.~\eqref{eq:four_terms}.

\begin{figure}[t]
\centering
\begin{tikzpicture}[
    mps/.style={circle,draw,fill=blue!25,minimum size=4mm,inner sep=0pt},
    zt/.style={rectangle,draw,fill=green!25,minimum size=4mm,inner sep=0pt},
    x=10mm,y=9mm]
  \def\NS{8}
  % wires
  \foreach \i in {1,...,8}{\draw (\i,0)--(\i,5.6);}
  % MPS
  \foreach \i in {1,...,8}{\node[mps] (m\i) at (\i,0) {};}
  \foreach \i in {1,...,7}{\pgfmathtruncatemacro{\j}{\i+1}\draw[thick] (m\i)--(m\j);}
  % even half gates (bonds 1-2,3-4,...)
  \foreach \i in {1,3,5,7}{\draw[fill=red!20] (\i-0.3,0.7) rectangle (\i+1.3,1.3);}
  % Z MPO half
  \foreach \i in {1,...,8}{\node[zt] (za\i) at (\i,2) {};}
  \foreach \i in {1,...,7}{\pgfmathtruncatemacro{\j}{\i+1}\draw[very thick,green!50!black] (za\i)--(za\j);}
  \foreach \i in {2,4,6}{\draw[fill=red!40] (\i-0.3,2.7) rectangle (\i+1.3,3.3);}
  % Z MPO half
  \foreach \i in {1,...,8}{\node[zt] (zb\i) at (\i,4) {};}
  \foreach \i in {1,...,7}{\pgfmathtruncatemacro{\j}{\i+1}\draw[very thick,green!50!black] (zb\i)--(zb\j);}
  \foreach \i in {1,3,5,7}{\draw[fill=red!20] (\i-0.3,4.7) rectangle (\i+1.3,5.3);}
  % labels
  \node[right,font=\small] at (8.6,1)   {$e^{-iH_{\rm e}\delta t/2}$};
  \node[right,font=\small] at (8.6,2)   {$e^{-iH_z\delta t/2}$};
  \node[right,font=\small] at (8.6,3)   {$e^{-iH_{\rm o}\delta t}$};
  \node[right,font=\small] at (8.6,4)   {$e^{-iH_z\delta t/2}$};
  \node[right,font=\small] at (8.6,5)   {$e^{-iH_{\rm e}\delta t/2}$};
  \node[left,font=\small] at (0.6,0) {$|\varphi(t)\rangle$};
  \draw[decorate,decoration={brace,amplitude=5pt}] (11.0,5.4)--(11.0,0.6)
     node[midway,right=6pt,font=\small,align=left] {$\times\,n_{\mathrm{step}}$\\ per segment $\Delta t$};
  \node[below,font=\small,align=center] at (4.5,-0.45) {Time evolution with bond dimension $\le\bdim$};
  \node[left,font=\scriptsize,green!40!black,align=right] at (0.65,2) {flux bond\\ $|L_n|\le\Lcut$};
\end{tikzpicture}
\caption{Schematic diagram for a single second-order Trotter step as in Eq.~\eqref{eq:trotter}. Red: two-site hopping gates on even and odd bonds. Green: exact MPO for the exponential of the diagonal part (mass and long-range electric energy), whose virtual index carries the electric flux $L_n$ on each link, truncated to $|L_n|\le\Lcut$ (bond dimension $2\Lcut+1$). }
\label{fig:tebd}
\end{figure}
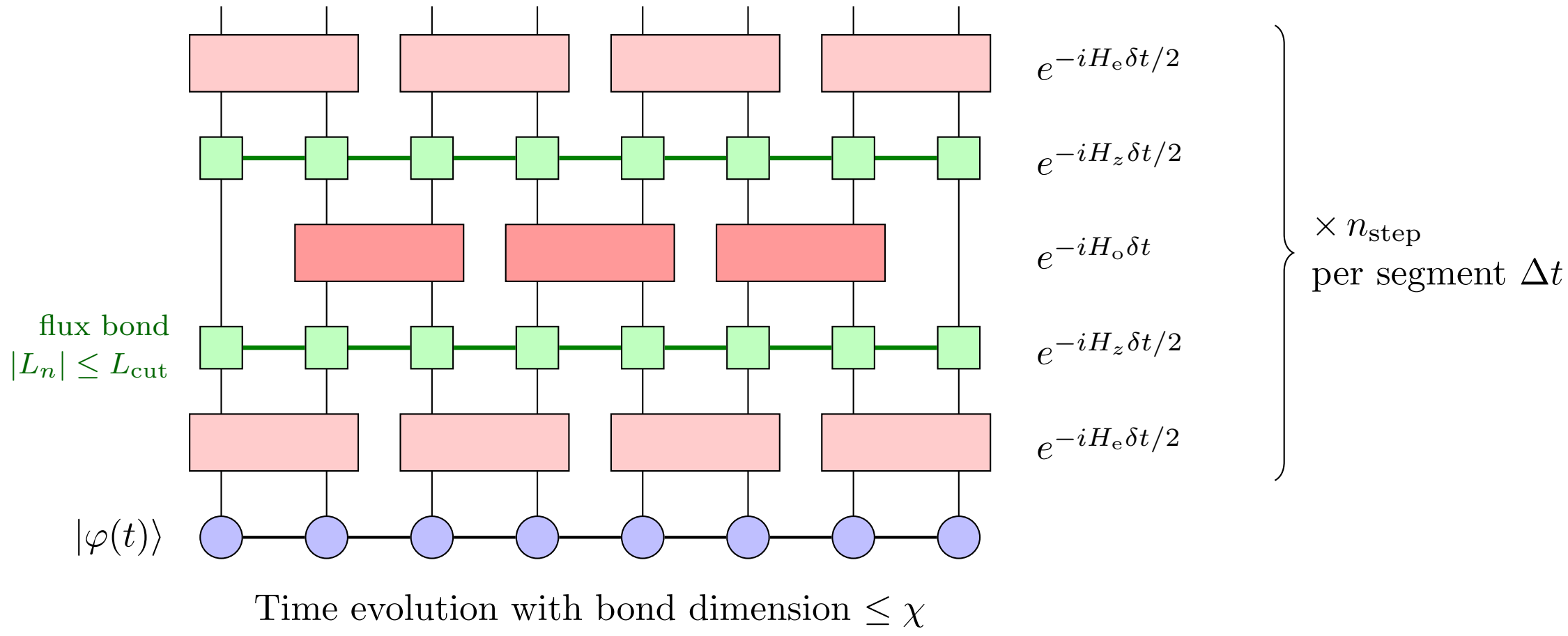

Importantly, the real-time evolution $e^{-iH_j\Delta t}$ is performed using the time-evolving block decimation (TEBD) algorithm~\cite{Vidal:2003lvx}. We split the Hamiltonian into hopping terms on even and odd bonds, $H_{\rm e}$ and $H_{\rm o}$, and the diagonal part $H_z$, which contains the mass term and the long-range electric energy including the static charge, and use the symmetric second-order Suzuki-Trotter decomposition
\begin{align}
    e^{-iH_j\delta t}\simeq
    e^{-iH_{\rm e}\delta t/2}\,e^{-iH_z\delta t/2}\,e^{-iH_{\rm o}\delta t}\,e^{-iH_z\delta t/2}\,e^{-iH_{\rm e}\delta t/2},
    \label{eq:trotter}
\end{align}
with $\delta t=\Delta t/n_{\mathrm{step}}$ and $n_{\mathrm{step}}=100$ substeps per segment. The hopping exponentials are products of commuting two-site gates. The electric term is non-local, but since it is diagonal in the $\sigma^z$ basis and $L_n$ obeys the recursion $L_n=L_{n-1}+[\sigma^z_n+(-1)^{n+1}]/2+q_n$, its exponential can be written exactly as an MPO whose virtual index carries the value of the electric flux on each link. Each MPO tensor maps the incoming flux $L_{n-1}$ and the local spin $\sigma^z_n$ to the outgoing flux $L_n$ and multiplies by the phase $\exp\{-i\delta t\,[\tfrac{m}{2}(1+(-1)^{n+1}\sigma^z_n)+\tfrac{ag^2}{2}L_n^2]\}$.
To reduce the computation complexity, we follow the recipe in Ref.~\cite{Banuls:2015sta,Banuls:2016lkq} to restrict the flux to $|L_n|\le\Lcut$ gives an MPO of bond dimension $2\Lcut+1$, which is exact for all configurations with smaller fluxes. An illustration is shown in Fig.~\ref{fig:tebd} for one Trotter step. After each layer the MPS is truncated to a maximal bond dimension $\bdim$ with cutoff $10^{-12}$. 

%% ============================================================================
\subsection{From correlators to distributions}
\label{sec:ft}
%% ============================================================================

On the lattice, the PDF follows from the discrete version of Eq.~\eqref{eq:pdf},
\begin{align}
    f_h(\xi)=\frac{M_hL}{4\pi}\sum_{d=-d_{\max}}^{d_{\max}}e^{-i\xi M_hz_d}\,\mathcal{M}_h(d)\,,
    \label{eq:pdf_lat}
\end{align}
where the sum runs over even $d$, with $M_h=E_h-E_0$, $z_d = d a$, and $L=Na$. The prefactor combines the relativistic normalization of the meson state with the measure of Eq.~\eqref{eq:pdf} and the normalization of the staggered fields. 

Analogously, the LCDA is obtained from the transition amplitude of Eq.~\eqref{eq:lcda},
\begin{align}
    R_h(u)=\frac{\sqrt{M_hL}}{2\pi}\sum_{d=-d_{\max}}^{d_{\max}}e^{iuM_hz_d}\,\mathcal{A}_h(d)\,,
    \label{eq:lcda_lat}
\end{align}
where the prefactor follows from the same steps as for the PDF. The sums can be evaluated on an arbitrarily fine grid in $\xi$ or $u$, but the resolution is limited by the range $|z|\le z_{d_{\max}}$ accessible on the lattice, and the results are periodic in $\xi$ and $u$ with period $2\pi/(M_h\delta z)=\pi/(M_ha)$. 

Following Eq.~\eqref{eq:fV} and Eq.~\eqref{eq:scalar_norm}, we normalize $\phi_V=R_V/\int_0^1R_V$ and $\phi_S=R_S/\int_0^1(2u-1)R_S$, and use $\mathsf{f}_V=|\int_0^1R_V|$ and $\mathsf{f}_S=|\int_0^1(2u-1)R_S|$. Equivalently, the decay constants can also be obtained from the local matrix element, $\mathsf{f}_V=\sqrt{M_VL}\,|\mathcal{A}_V(0)|/(2aM_V)$ and 
% $\mathsf{f}_S=\sqrt{M_SL}\,|d\mathcal{A}/dz|_{z=0}/(aM_S^2)$.
$\mathsf{f}_S=\sqrt{M_SL}\,\big|\tfrac{d}{dz}\big[e^{iM_Sz/2}\mathcal{A}_S(z)\big]_{z=0
}\big|/(aM_S^2)$.
Both the PDF and the normalized LCDA distributions should be real; a residual imaginary part provides an estimate of lattice artifacts. 

%% ============================================================================
\section{Numerical results}
\label{sec:results}
%% ============================================================================

In this work, we focus on lattices of $N=100$ sites with $ga=0.1$ ($x=100$) and physical volume $\mathcal{V}=10$. We mainly consider four fermion masses  $m/g=1/\sqrt{\pi}$, $5/\sqrt{\pi}$, $10/\sqrt{\pi}$, and $20/\sqrt{\pi}$ ($\approx0.564$, $2.82$, $5.64$, $11.28$) with our setup. The scalar and vector meson states are obtained with DMRG with maximum bond dimension $D=400$, and their light-cone correlators for the PDFs and LCDAs are computed with TEBD at maximum bond dimension $\bdim=64$ and maximum flux truncation $\Lcut=16$ for $|d|\le48$. Convergence studies of higher bond dimensions, flux truncation, and lattice spacing for the TEBD is included in App.~\ref{app:convergence}.

%% ============================================================================
\subsection{Meson identification}
\label{sec:spectroscopy}
%% ============================================================================

\begin{figure}[t]
    \centering
    \includegraphics[width=0.48\linewidth]{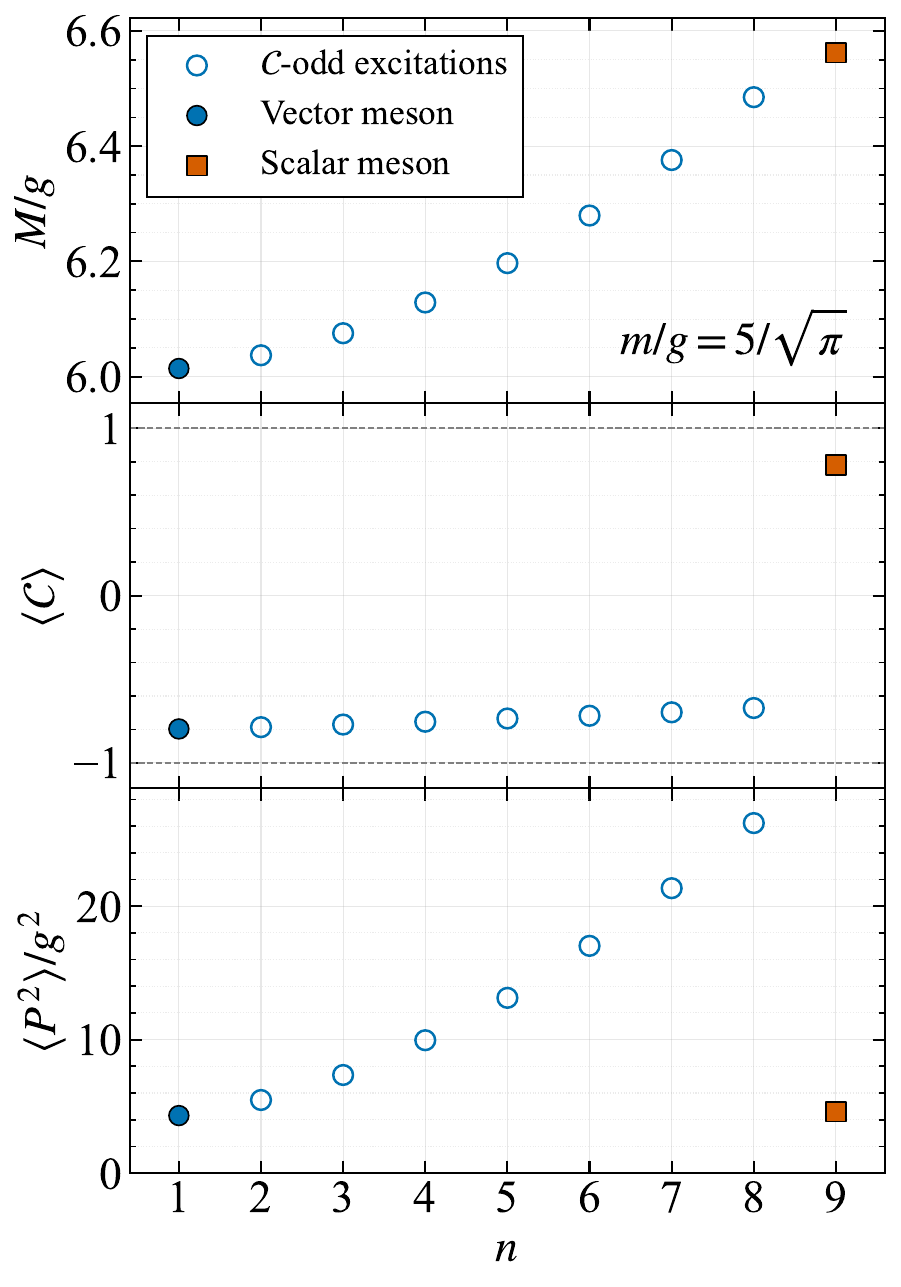}\,
    \includegraphics[width=0.49\linewidth]{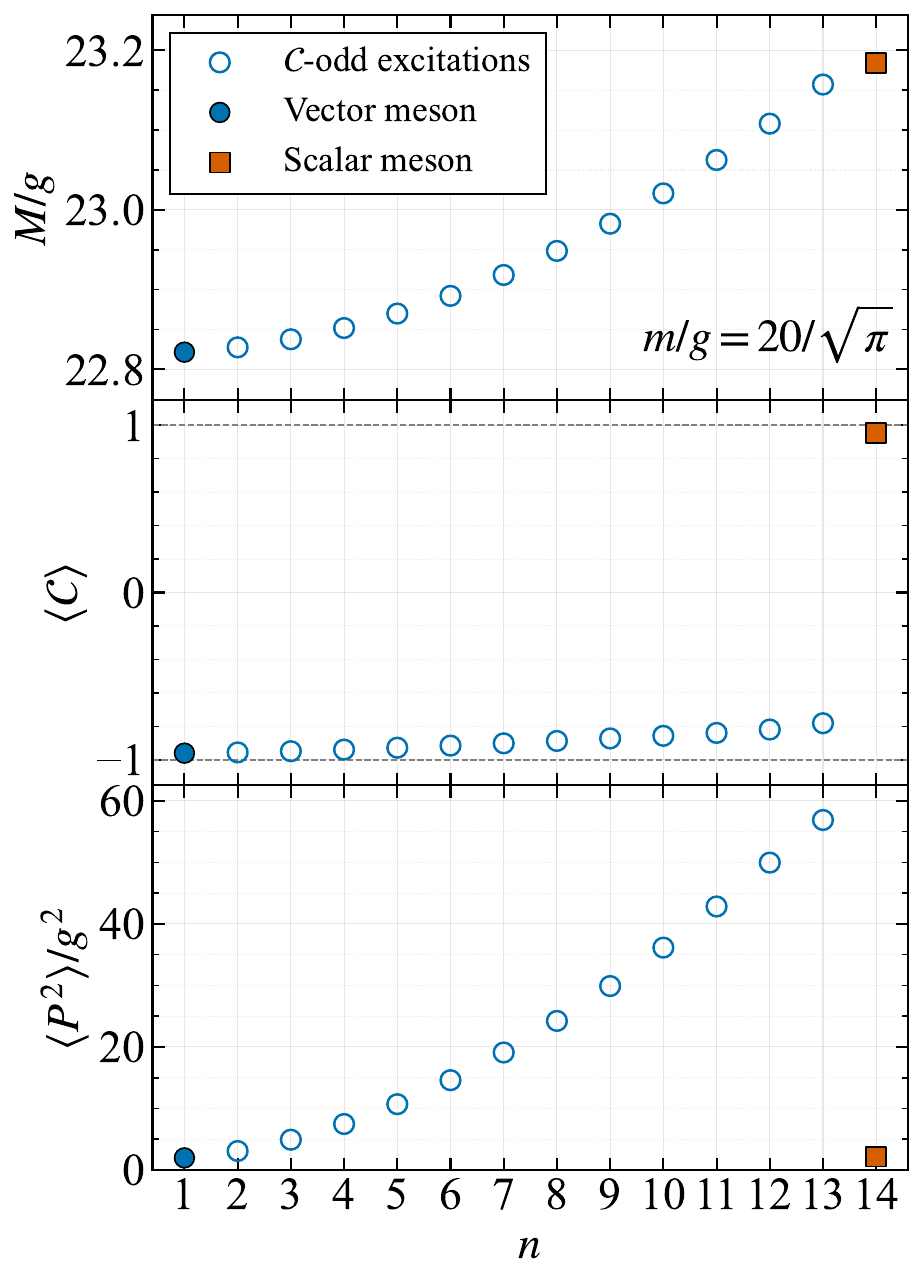}
    \caption{State identification for $N=100$, $ga=0.1$, at $m/g=5/\sqrt{\pi}$ (left) and $20/\sqrt{\pi}$ (right). From top to bottom: mass gap $M/g$, lattice charge conjugation $\langle\Cop\rangle$, and squared lattice momentum $\langle\hat P^2\rangle/g^2$, for the lowest states of the zero-charge sector. Open circles: $\Cc$-odd vector-meson excitations; filled circle: vector meson near at rest; square: scalar meson. 
    }
    \label{fig:spectrum}
\end{figure}

\begin{table}[t]
    \centering
    \begin{tabular}{c|ccccc|ccc}
    \hline
     & \multicolumn{5}{c|}{Tensor network result} & \multicolumn{3}{c}{Non-relativistic model } \\
    \cline{2-9}
    $m/g$ & $\langle\mathcal{C}\rangle_V$ & $\langle\mathcal{C}\rangle_S$ & $M_V/g$ & $M_S/g$ & $(M_S-M_V)/g$ & $M_V/g$ & $M_S/g$ & $(M_S-M_V)/g$ \\
    \hline
    $1/\sqrt{\pi}$  & $-0.55$ & $0.55$ & $1.64$ & $2.35$ & $0.71$ & $1.91$  & $2.91$  & $1.01$ \\
    $5/\sqrt{\pi}$  & $-0.80$ & $0.78$ & $6.01$  & $6.56$  & $0.55$ & $6.10$  & $6.68$  & $0.59$ \\
    $10/\sqrt{\pi}$ & $-0.96$ & $0.95$ & $11.60$ & $12.05$ & $0.45$ & $11.64$ & $12.11$ & $0.47$ \\
    $20/\sqrt{\pi}$ & $-0.90$ & $0.87$ & $22.82$ & $23.18$ & $0.36$ & $22.85$ & $23.22$ & $0.37$ \\
    \hline
    \end{tabular}
    \caption{Charge-conjugation expectation values and masses of the vector and scalar mesons from the tensor-network simulation with $N=100$ and $ga=0.1$, compared with the non-relativistic model in Eq.~\eqref{eq:hf_mass}.}
    \label{tab:masses}
\end{table}

We first compute the low-lying spectrum of the zero-charge sector with DMRG and identify the states through $\langle\Cop\rangle$ and $\langle P^2\rangle$. As an example, we show in Fig.~\ref{fig:spectrum}, the lowest states for $m/g=5/\sqrt{\pi}$ and $20/\sqrt{\pi}$. For $m/g=5/\sqrt{\pi}$, the states $n=1,2,\dots,8$ have $\langle\Cop\rangle<0$ and form a single $\Cc$-odd branch, along which $\langle P^2\rangle_n$ and the energy increase monotonically, as expected for the vector meson in the standing-wave momentum modes $p_n=n\pi/L$ of the open box with box size $L=Na$. Thus, $n_V=1$ is the vector meson near at rest, and the higher states are the vector meson with larger momenta. The state $n=9$ breaks this pattern, with $\langle\Cop\rangle\simeq+0.8$ and $\langle P^2\rangle$ close to that of $n=1$, and we identify it as the scalar meson near at rest, so $n_S=9$. The same procedure is used to find the scalar meson for all the different masses. For $m/g=1/\sqrt{\pi}$, not shown in the figure, it appears at $n_S=6$, but with $|\langle\Cop\rangle|\simeq0.55$ for both mesons, indicating sizable finite-volume and lattice effects at this mass. We show the respective meson mass $M_V$, $M_S$, and their mass gap $M_S-M_V$, along with the $\langle\Cop\rangle$ in Table~\ref{tab:masses}; the deviation $1-|\langle\Cop\rangle|$ measures the breaking of charge conjugation by the open boundaries.

The position of the scalar meson can be estimated from the box modes of the vector meson, $E_n=(M_0^2+n^2\pi^2/L^2)^{1/2}$ with $E_1=M_V$. The vector states with $n\le K$ lie below the scalar meson, where $E_K=M_S$ gives
\begin{align}
    K=\Big[1+\frac{(M_S^2-M_V^2)L^2}{\pi^2}\Big]^{1/2}\,,
    \label{eq:K}
\end{align}
so that the scalar meson is expected at $n_S=\lfloor K\rfloor+1$. From the masses obtained in the tensor network, Eq.~\eqref{eq:K} gives $K=5.4$, $8.4$, $10.4$, and $13.0$ for the four masses, consistent with the observed $n_S=6$, $9$, $11$, and $14$. In the non-relativistic limit, note that the number of vector states below the scalar meson grows with $m/g$, because the spacing of the box modes goes with $1/(M_VL^2)\approx 1/m$, which decreases faster than the vector-scalar splitting that grows with $m^{-1/3}$ as in Eq.~\eqref{eq:hf_mass}. In Table~\ref{tab:masses}, the tensor-network results agree with the NR model within a few percent for $m/g\ge5/\sqrt{\pi}$, with the agreement improving as the mass increases; however, at $m/g=1/\sqrt{\pi}$ they deviate significantly, which is as expected for $m\lesssim g$. Instead, the vector mass is in good agreement with the second-order small-mass expansion~\cite{Sriganesh:1999ws,Adam:1997wt,Vary:1996uc},
\begin{align}
    \frac{M_V}{g}\simeq\frac{1}{\sqrt\pi}+e^{\gamma_E}\frac{m}{g}+0.1907\Big(\frac{m}{g}\Big)^2\,,
    \label{eq:mpt}
\end{align}
with $\gamma_E\simeq 0.577$ the Euler constant, which gives $M_V/g\simeq1.63$ at $m/g=1/\sqrt{\pi}$.

%% ============================================================================
\subsection{Parton structure for vector meson}
\label{sec:vector}
%% ============================================================================

\begin{figure}[t]
    \centering
    \includegraphics[width=0.47\linewidth]{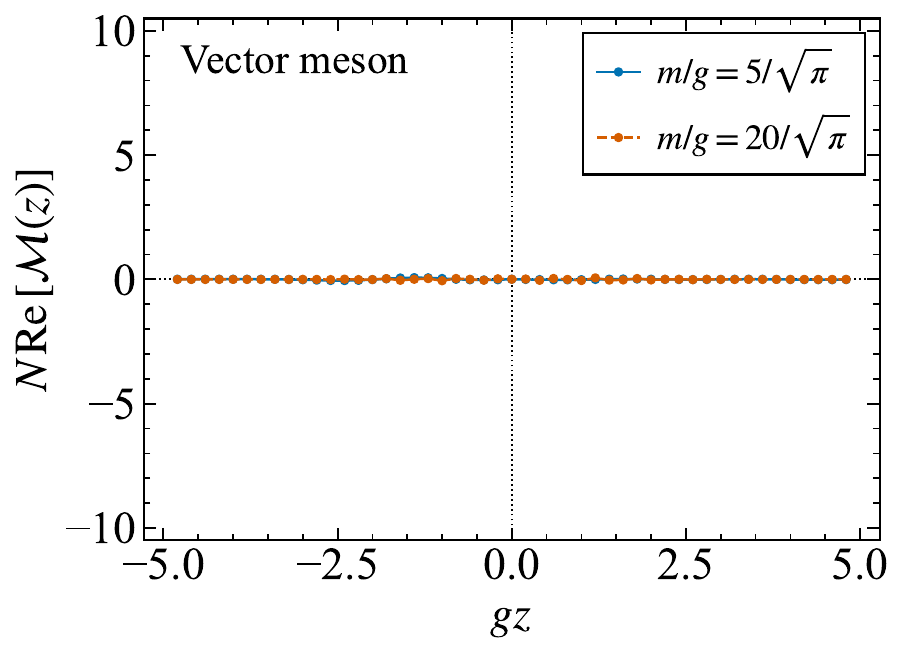}
    \hfill
    \includegraphics[width=0.47\linewidth]{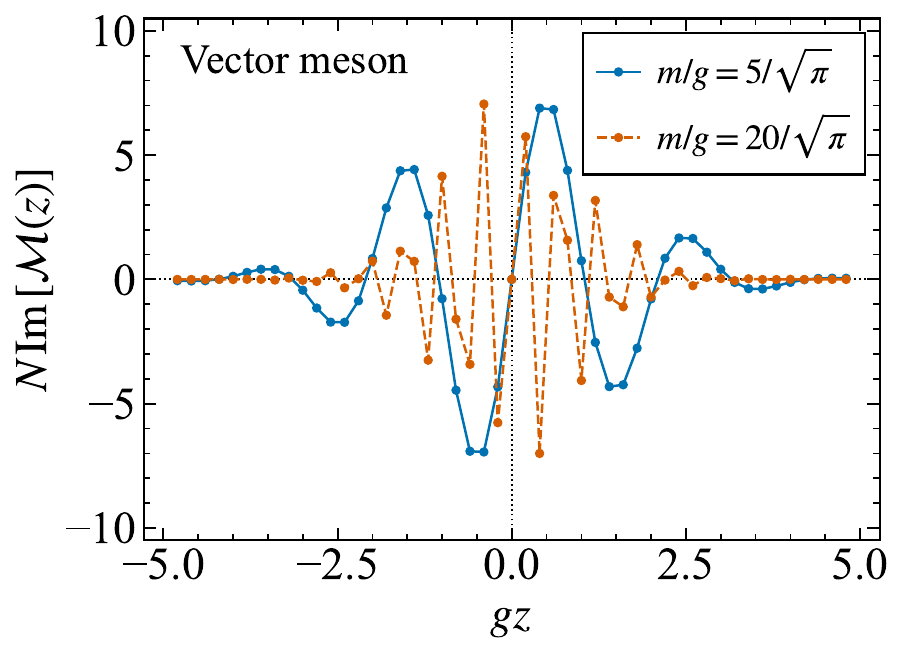}
    \includegraphics[width=0.47\linewidth]{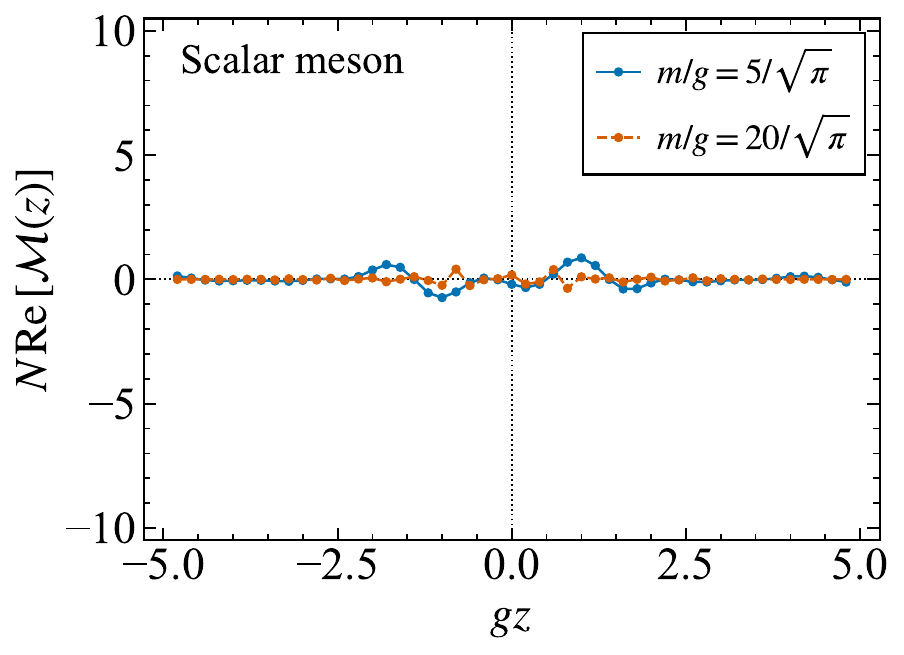}
    \hfill
    \includegraphics[width=0.47\linewidth]{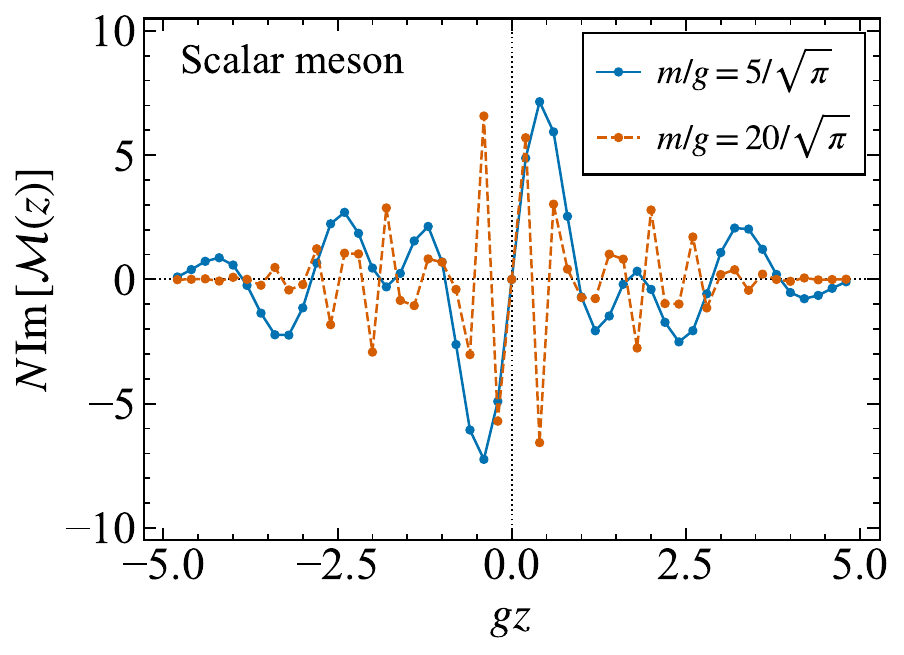}
    \caption{Real (left) and imaginary (right) parts of the vacuum-subtracted PDF correlator $N\mathcal{M}(z)$ of the vector meson and scalar meson versus the light-cone separation $gz$ for $m/g=5/\sqrt{\pi}$ and $20/\sqrt{\pi}$. }
    \label{fig:me_pdf}
\end{figure}

\begin{figure}[t]
    \centering
    \includegraphics[width=0.47\linewidth]{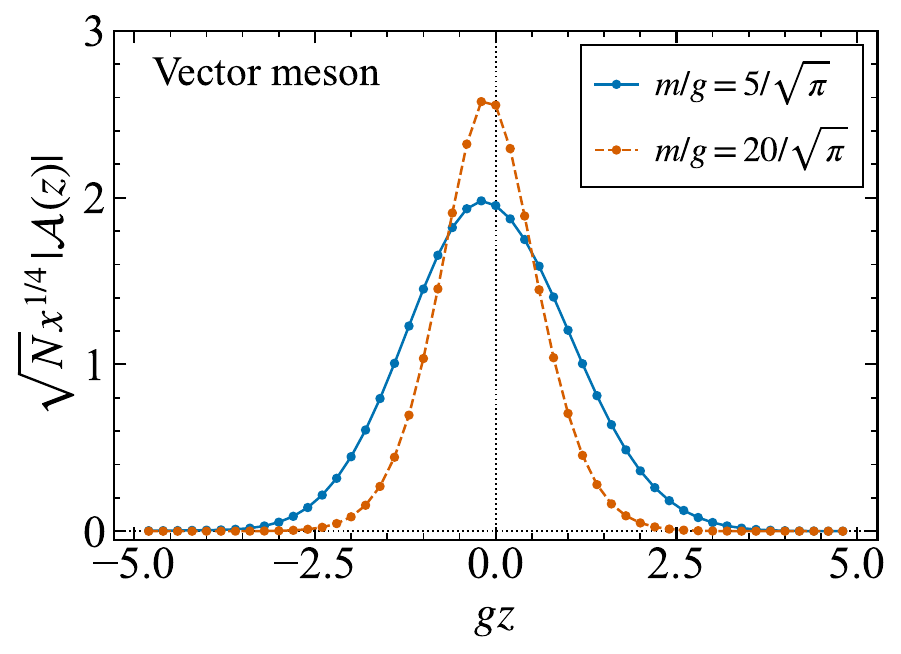}
    \hfill
    \includegraphics[width=0.47\linewidth]{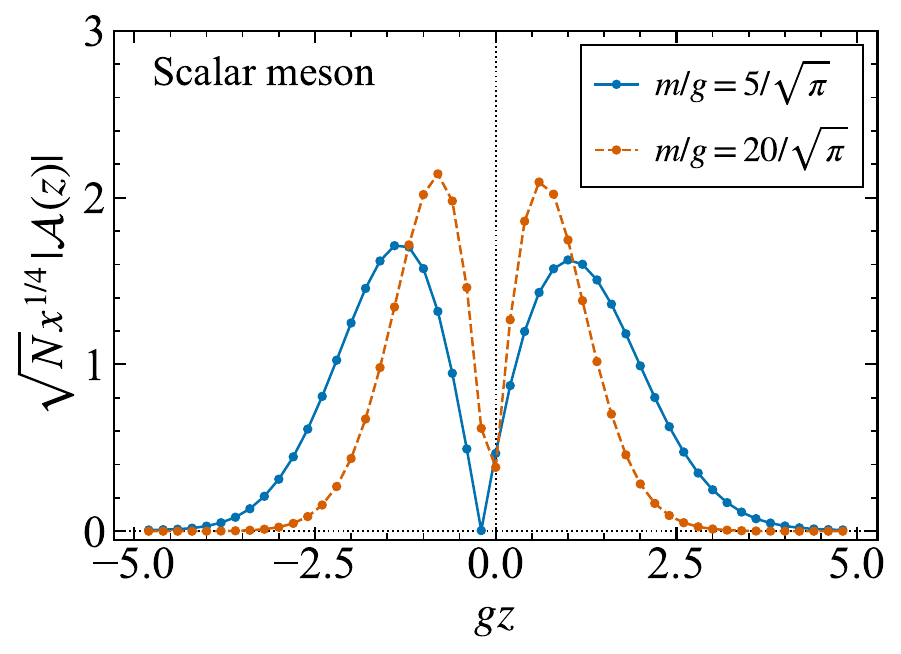}
    \caption{The LCDA correlator $\sqrt{N}|\mathcal{A}(z)|$ of the vector meson (left) and the scalar meson (right) versus the light-cone separation $gz$ for $m/g=5/\sqrt{\pi}$ and $20/\sqrt{\pi}$.}
    \label{fig:me_A}
\end{figure}

\begin{table}[t]
    \centering
    \begin{tabular}{c|ccc|cccc}
    \hline
     & \multicolumn{3}{c|}{$\mathsf{f}_V$} & \multicolumn{4}{c}{$\mathsf{f}_S$} \\
    \cline{2-8}
    $m/g$ & local  & $|\int_0^1du\,R_V|$ & NR model & local & $|\int_0^1du\,(2u-1)\,R_S|$ & NR model & $\epsilon_S$ \\
    \hline
    $1/\sqrt{\pi}$  & $0.520$ & $0.497$ & $0.755$ & $0.302$ & $0.266$ & $0.886$ & 0.028\\
    $5/\sqrt{\pi}$  & $0.398$ & $0.399$ & $0.442$ & $0.131$ & $0.133$ & $0.177$ & 0.087\\
    $10/\sqrt{\pi}$ & $0.331$ & $0.331$ & $0.351$ & $0.082$ & $0.084$ & $0.089$ & 0.010\\
    $20/\sqrt{\pi}$ & $0.267$ & $0.268$ & $0.278$ & $0.042$ & $0.044$ & $0.044$ & 0.040\\
    \hline
    \end{tabular}
    \caption{Decay constants of the vector and scalar mesons for $N=100$ and $ga=0.1$ computed using local matrix elements at $z=0$ and moment normalization, compared with non-relativistic (NR) model. Note $\epsilon_S = |\int_0^1du\,R_S|$ is used to measure lattice effect for the scalar meson, and $\mathsf{f}_V=1/\sqrt{\pi}\simeq0.564$ in the massless limit, }
    \label{tab:decay}
\end{table}

\begin{figure}[t]
    \centering
    \includegraphics[width=0.47\linewidth]{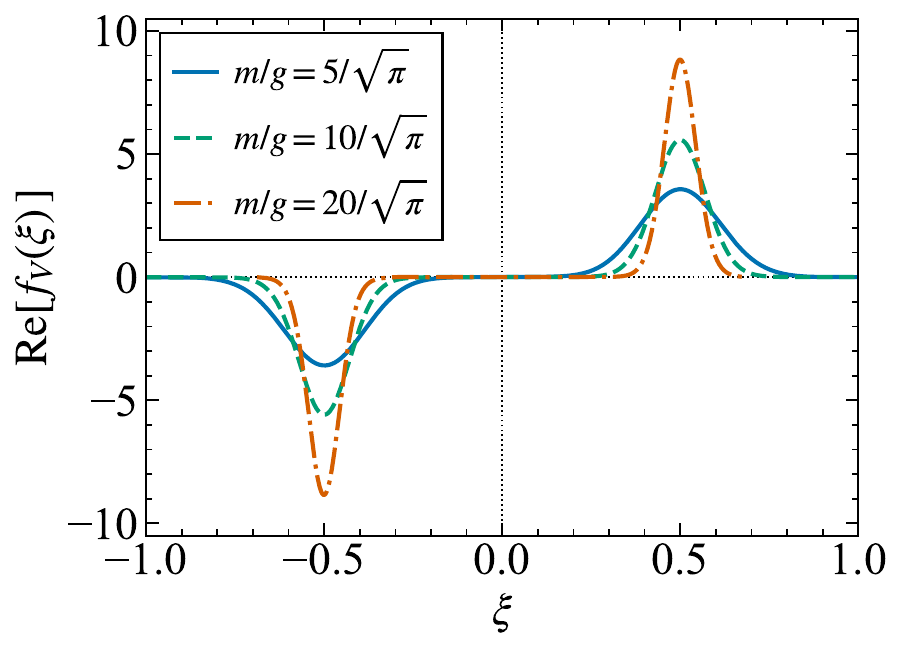}
    \hfill
    \includegraphics[width=0.47\linewidth]{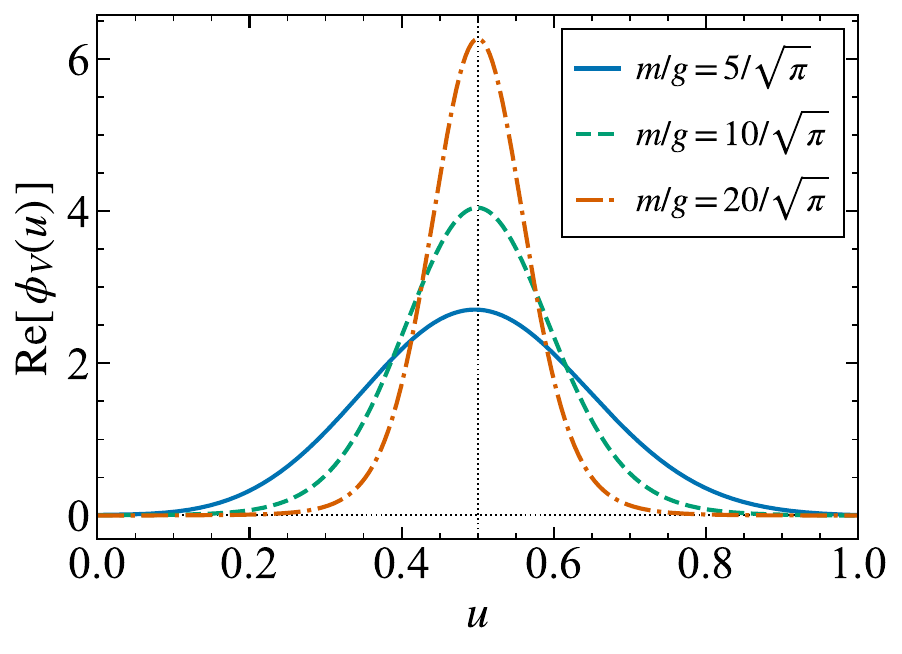}
    \caption{PDF $f_V(\xi)$ (left) and LCDA $\phi_V(u)$ (right) of the vector meson for $m/g=5/\sqrt\pi$, $10/\sqrt\pi$, and $20/\sqrt\pi$.}
    \label{fig:PDF_and_LCDA_vector}
\end{figure}

We first examine the parton structures of the vector mesons identified as the first excitation with $\langle\Cop\rangle\simeq -1$ in our spectrum. Figure~\ref{fig:me_pdf} shows the vacuum-subtracted PDF correlator in Eq.~\eqref{eq:MA} of the vector meson before performing the Fourier transform for two selected mass values. For a $\Cc$ eigenstate of definite momentum as well as the vacuum, the hermiticity of $\mathcal{O}$ gives $\mathcal{M}_h(-z)=\mathcal{M}_h(z)^*$, while charge conjugation gives $\mathcal{M}_h(-z)=-\mathcal{M}_h(z)$. Therefore, $\mathcal{M}_h$ is purely imaginary and odd in $z$, which makes $f_h$ real and odd in $\xi$. Moving onto $\mathcal{A}_h(z)$, the matrix elements for LCDA, charge conjugation exchanges the endpoints of $\mathcal{O}$, and translating back yields a phase from the meson momentum, $\mathcal{A}_h(z)=-\eta_{\Cc}\,e^{-iM_hz}\mathcal{A}_h(-z)$. So, $|\mathcal{A}_h(z)|$ is expected to be even in $z$ for both the vector and scalar mesons, and its numerical results at selected masses are presented in Fig.~\ref{fig:me_A}. The correlators in Figs.~\ref{fig:me_pdf} and~\ref{fig:me_A} satisfy these properties to good accuracy, even though translation invariance and charge conjugation are not exact symmetries on a finite lattice with open boundary conditions and staggered fermions. The residual violations are most visible in $|\mathcal{A}_h(z)|$ at the coarsest lattice spacing, where its center is displaced slightly toward negative $z$, by less than one lattice spacing, and the two peaks of the scalar meson have slightly different heights. Both effects decrease as the lattice spacing is reduced at fixed physical volume in Fig.~\ref{fig:conv_fixed_volume} (in App.~\ref{app:convergence}), indicating them as lattice artifacts. We include the prefactors $N$ and $\sqrt{N}\,x^{1/4}$ in $\mathcal{M}_h$ and $\mathcal{A}_h$, respectively, which absorb the normalization of the lattice fields and of the unit-normalized states, so that the results can be compared directly across lattice spacings.

Now we proceed to compute the PDF and the LCDA of the vector meson using discrete Fourier transform of these matrix elements. Figure~\ref{fig:PDF_and_LCDA_vector} shows the PDF and the LCDA of the vector meson for the three different masses. Here, we present only the real part of the PDF and the LCDA. It is clear that our PDF is odd in $\xi$, with a peak at $\xi\simeq1/2$ and its mirror image at $\xi\simeq-1/2$. For $m/g=5/\sqrt\pi$ and $10/\sqrt\pi$, the PDF is negligible for $|\xi|>1$, although this is not imposed by the method; at $m/g=20/\sqrt\pi$, the Fourier period $\pi/(M_Va)\simeq1.38$ restricts the accessible range to $|\xi|\lesssim0.69$ though it is sufficient to obtain the shape. As expected, the PDF narrows with increasing mass toward $\xi=\pm1/2$. 
On the other hand, the LCDA is symmetric about $u=1/2$ as expected, and narrows with the same scaling. After division by the complex normalization, a small imaginary part remains, whose relative size grows with the mass. Note our PDF results are in agreement with previous calculations in Refs.~\cite{Banuls:2024oxa,Banuls:2025wiq}.  

Table~\ref{tab:decay} presents the results of the decay constants $\mathsf{f}_V$ for vector mesons calculated in two independent ways, either from the local matrix elements, i.e. the value and slope of the correlator at $z=0$, or from the normalization moments of the reconstructed LCDAs. The former require no Fourier transform, so their agreement tests the reconstruction from the finite range $|z|\le z_{\max}$. For $m/g\ge5/\sqrt{\pi}$, the two method agree exceptionally well for both mesons; at $m/g=1/\sqrt{\pi}$ they differ by $5\%$ for $\mathsf{f}_V$ and $14\%$ for $\mathsf{f}_S$, where the slower decay of the correlator makes the truncation more important. Overall, both decay constants decrease with the mass. 
Compared to the NR model, the tensor-network results mostly agree for $m/g\ge5/\sqrt{\pi}$, and the suppression of $\mathsf{f}_S/\mathsf{f}_V\propto(g/m)^{2/3}$ from Eq.~\eqref{eq:hf_decay} at higher $m/g$ values is observed in the table.

%% ============================================================================
\subsection{Parton structure for scalar meson}
\label{sec:scalar}
%% ============================================================================

We now turn to the scalar meson, which are identified as the lowest excitations with $\langle\Cop\rangle\simeq +1$. Unlike the vector meson, the scalar correlator $\mathcal{M}_S$, shown in Fig.~\ref{fig:me_pdf}, has a sizable real part, which is odd in $z$. Charge conjugation alone requires $\mathcal{M}_h$ to be purely imaginary, so this real part signals residual $\Cc$ breaking in the numerical scalar state. Consistently, the effect is largest at the smaller mass, $m/g=5/\sqrt\pi$, where the deviation $1-\langle\Cop\rangle_S$ in Table~\ref{tab:masses} is also largest. It is not caused by insufficient convergence of the bound state as the energy variance of the scalar state, $\langle H^2\rangle-\langle H\rangle^2$, is below $10^{-7}g^2$. Instead, it reflects the breaking of charge conjugation by the open boundaries and the staggered discretization, which affects the highly excited scalar state more strongly than the vector meson.

\begin{figure}[t]
    \centering
    \includegraphics[width=0.47\linewidth]{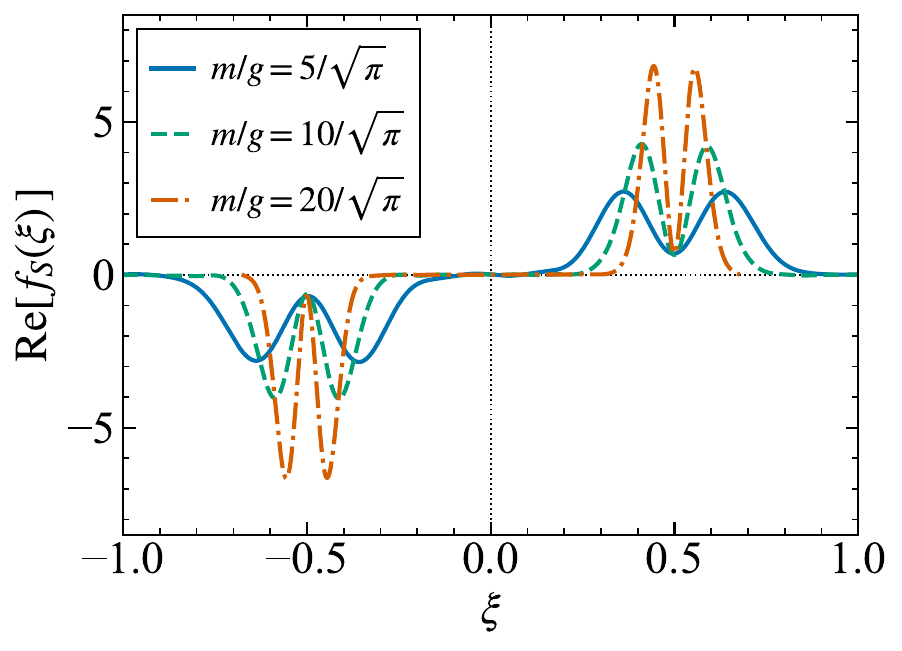}
    \hfill
    \includegraphics[width=0.47\linewidth]{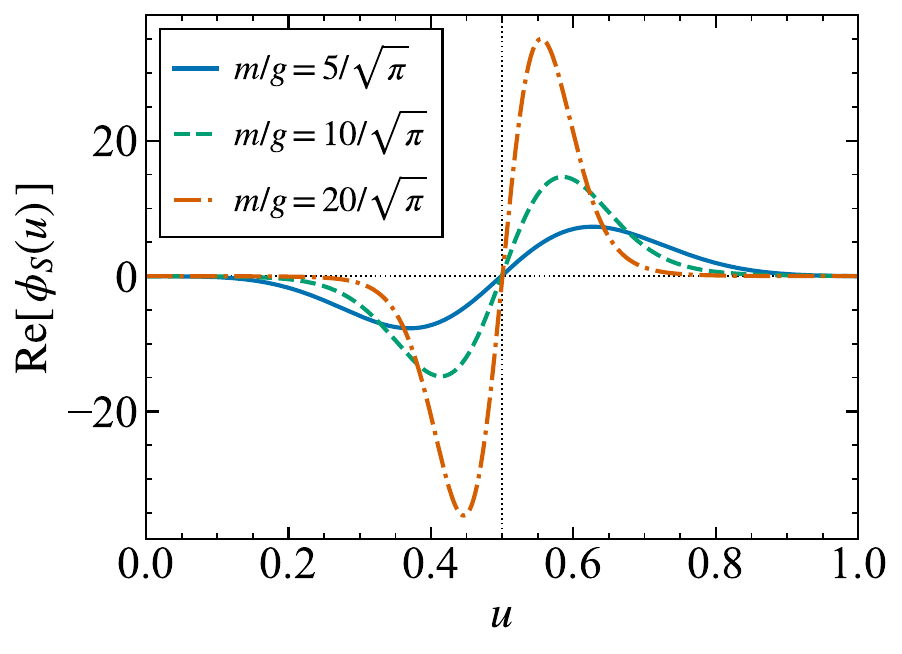}
    \caption{PDF $f_S(\xi)$ (left) and LCDA $\phi_S(u)$ (right) of the scalar meson for $m/g=5/\sqrt\pi$, $10/\sqrt\pi$, and $20/\sqrt\pi$. 
    \label{fig:PDF_and_LCDA_scalar}}
\end{figure}

The PDF and the LCDA for the scalar meson are presented in Fig.~\ref{fig:PDF_and_LCDA_scalar}. The PDF has the same symmetry properties as that of the vector meson, but a qualitatively different shape: two peaks located symmetrically about $\xi=1/2$, separated by a pronounced minimum. As the mass increases, the peaks approach each other as expected, from the narrowing of the momentum-space wave function in the non-relativistic limit, Eq.~\eqref{eq:hf_dist}. The minimum is the momentum-space image of the node of the odd relative wavefunction. In the valence approximation, $f_S(1/2)\propto|\phi_S(1/2)|^2=0$. The observed minimum does not reach zero, which could be related to either higher Fock sector contribution or finite-volume effects that was not studied systematically in our calculations.

The LCDA is antisymmetric about $u=1/2$, with a node at $u=1/2$ and two peaks of opposite sign that approach each other and sharpen as the mass increases. With the first-moment normalization the peaks grow rapidly with the mass, since a narrower distribution needs a larger amplitude to produce a unit first moment. The zeroth moment of $R_S$, which vanishes for an exact $\Cc$-even state, measures the residual charge-conjugation breaking of the scalar states and is listed as $\epsilon_S$ in Table~\ref{tab:decay}.

The node structure of the LCDA is a direct consequence of charge conjugation. Since $\phi_S(1-u)=-\phi_S(u)$, the valence amplitude vanishes when the fermion and the antifermion carry equal momentum fractions, so that two constituents share the momentum unequally, with either constituent carrying the larger fraction. In the valence approximation, this node also produces the minimum of the PDF. In the non-relativistic picture, it corresponds to the node at $r=0$ of the odd relative wave function, whose peaks lie at $|p|\sim1/r_0$, so $|u-\tfrac12|\sim1/(2mr_0)\propto(g/m)^{2/3}$, therefore approaching each other as the mass increases. The same antisymmetric structure appears for the LCDAs of $P$-wave scalar quarkonia~\cite{Braguta:2008qe,Li:2026fgc}.

%% ============================================================================
\subsection{PDFs versus squared LCDAs}
\label{sec:comparison}
%% ============================================================================

\begin{figure}[t]
    \centering
    \includegraphics[width=0.47\linewidth]{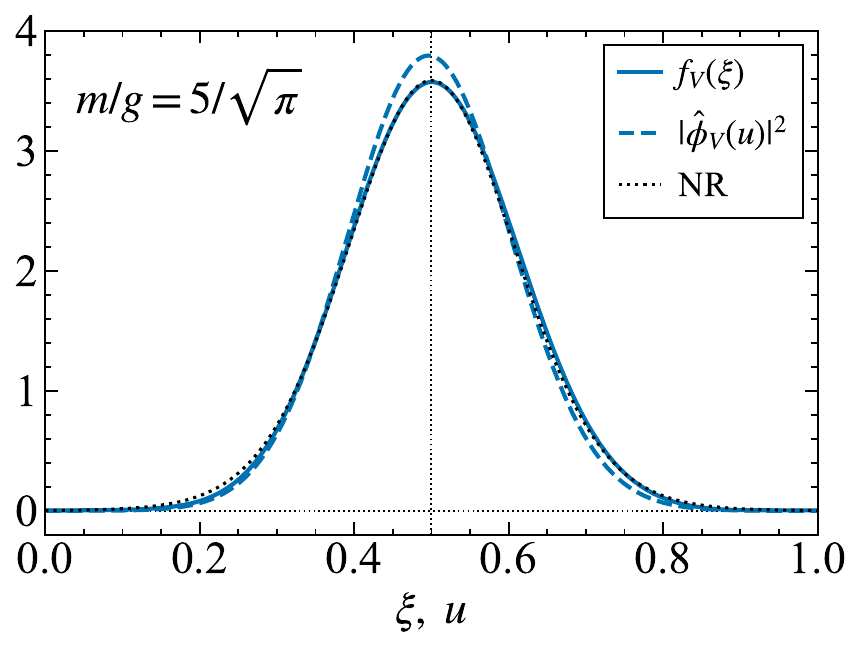}
    \hfill
    \includegraphics[width=0.47\linewidth]{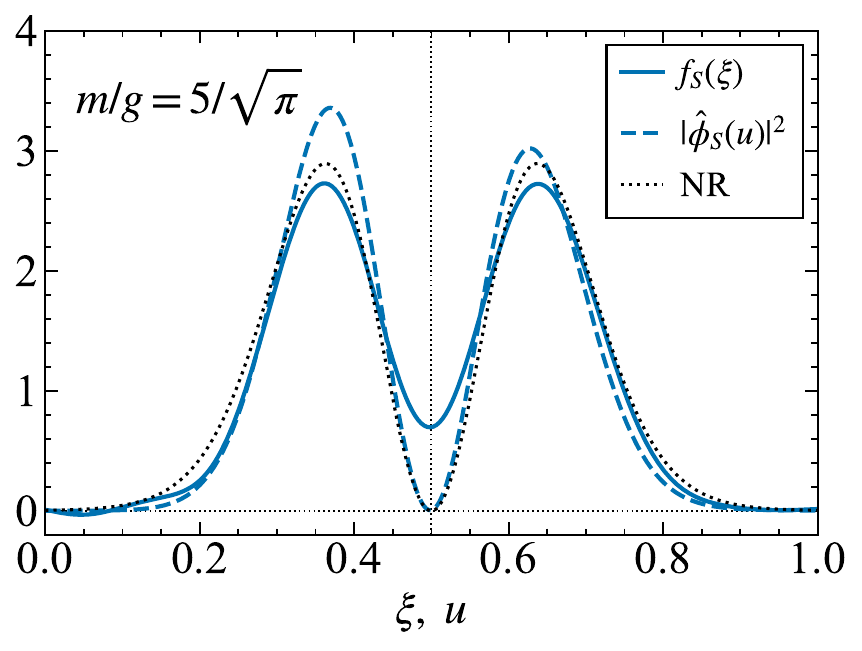}
    \includegraphics[width=0.47\linewidth]{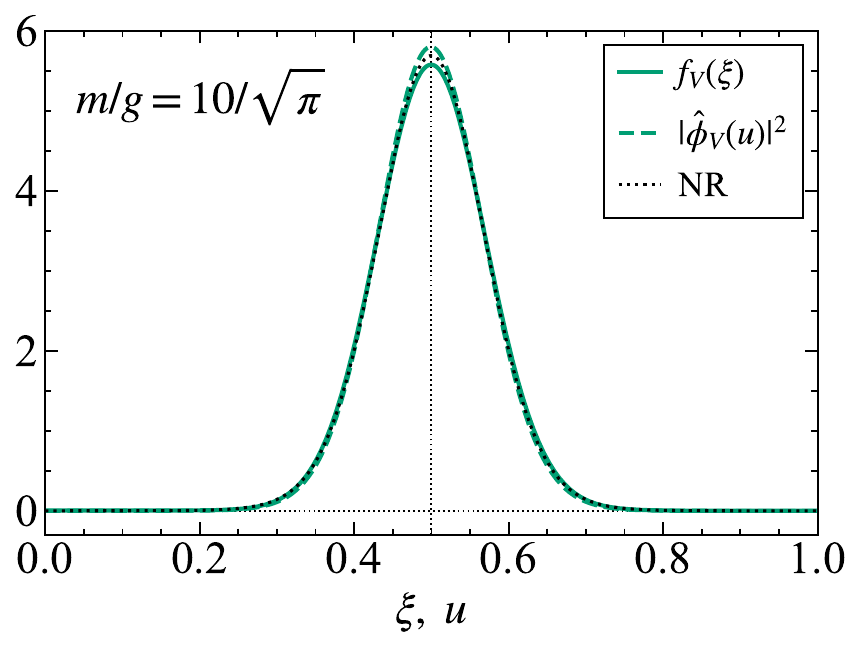}
    \hfill
    \includegraphics[width=0.47\linewidth]{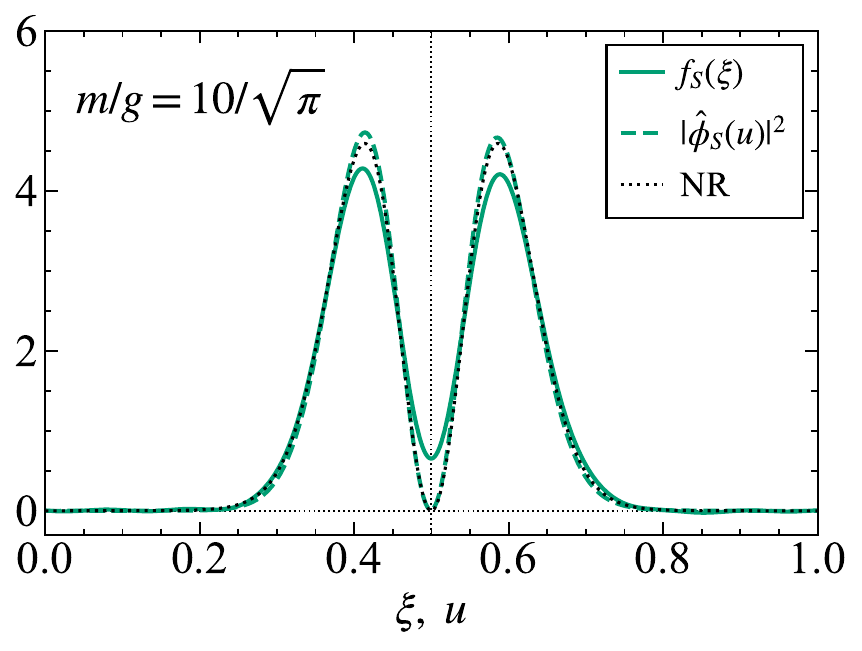}
    \includegraphics[width=0.47\linewidth]{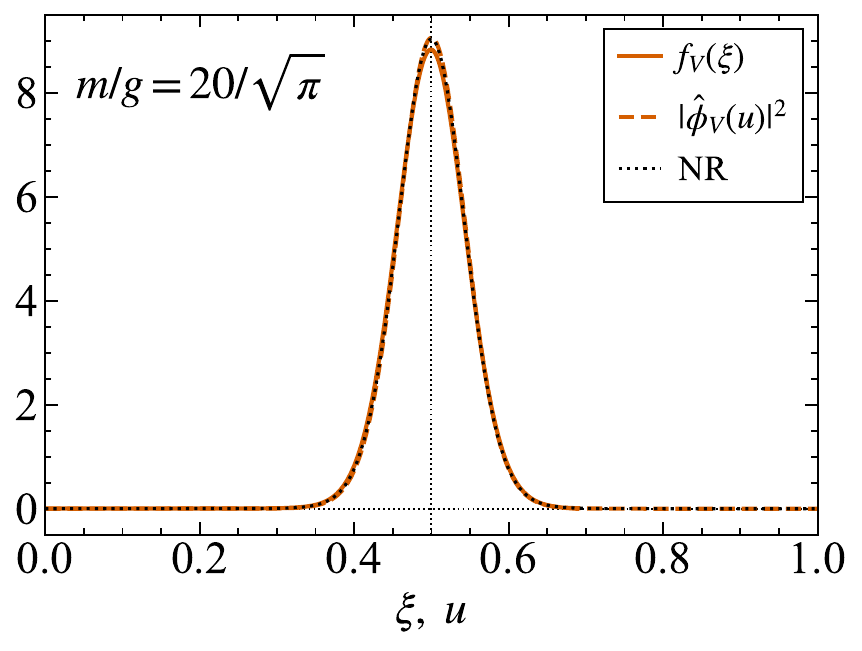}
    \hfill
    \includegraphics[width=0.47\linewidth]{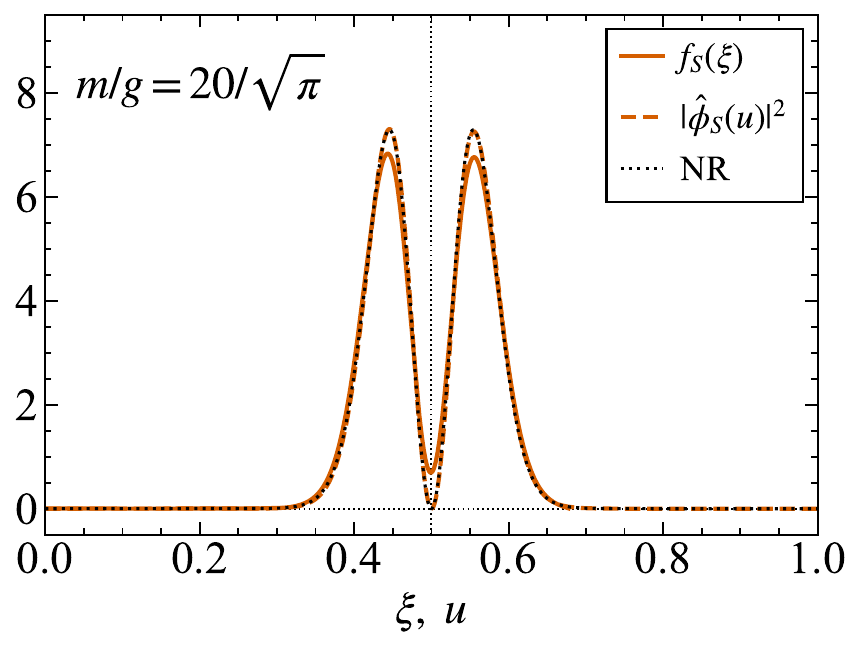}
    \caption{PDF $f_h(\xi)$ (solid) compared with the unit-normalized squared LCDA $|\hat\phi_h(u)|^2$ (dashed) and the NR model (dotted) for the vector (left) and scalar (right) mesons at $m/g=5/\sqrt\pi$, $10/\sqrt\pi$, and $20/\sqrt\pi$.}
    \label{fig:PDF_vs_LCDA2}
\end{figure}

Having both distributions for each meson, we compare in Fig.~\ref{fig:PDF_vs_LCDA2} the PDF with the unit-normalized squared LCDA $|\hat\phi_h|^2\equiv|\phi_h(u)|^2/\int_0^1du\,|\phi_h|^2$, together with the NR model. For $m/g=10/\sqrt\pi$ and $20/\sqrt\pi$, the LCDAs $|\hat\phi_h|^2$ agrees closely with the NR model for both mesons, confirming that the valence structure is well described by a linear potential; For $m/g=5/\sqrt\pi$, the deviations from the model become visible. For the reference, the momentum sum rules is well satisfied for $m/g = 5/\sqrt{\pi}, 10/\sqrt{\pi}, 20/\sqrt{\pi}$, with $\langle \xi \rangle_V \simeq0.503, 0.500, 0.499$ for the vector mesons, and $\langle \xi\rangle_S \simeq 0.505, 0.500, 4.99$ for the scalar ones.

For a pure fermion-antifermion state, $f_h$ and $|\hat\phi_h|^2$ are expected to coincide exactly in 1+1 dimensions, since there is no transverse momentum to integrate over. Therefore, their difference is sensitive to contributions from higher Fock components beyond the valence sector, which enter the PDF but not the LCDA. At $m/g=5/\sqrt\pi$, the PDF follows the shape of $|\hat\phi_h|^2$ closely, but is slightly broader and lower at the peaks, and $f_S$ remains finite at $\xi=1/2$, where $|\hat\phi_S|^2$ vanishes. Such differences could be a qualitative signature of higher Fock components, which reduce the valence contribution and add a broad multi-particle contribution. However, at the present precision, these differences may not be be attributed to higher Fock components alone, since lattice artifacts affect the two distributions differently. The PDF is a density and the LCDA an amplitude. Any smearing symmetric under $u\to1-u$, such as the truncation of the correlators at $z_{\max}$ or the center-of-mass motion of the meson in the open box, preserves the node of $\phi_S$ but fills that of $f_S$, and the admixture of $\Cc$-odd states allowed by the open boundaries also complicates the contribution at $\xi=1/2$. Separating these effects requires a more comprehensive and systematic study of the dependence on $z_{\max}$ and on the volume, which we leave for future work.

%% ============================================================================
\section{Conclusion and outlook}
\label{sec:conclusion}
%% ============================================================================
In this work, we have computed the PDFs and LCDAs of the vector and scalar mesons of the single-flavor massive Schwinger model from first principles with tensor networks. The meson states were prepared as MPS with DMRG and identified among the low-lying states through their corresponding charge-conjugation and momentum. The light-cone correlators were evaluated directly in Minkowski space by real-time evolution along a discretized light-cone path, with the Wilson line implemented through static charges, and the electric energy exponentiated exactly through an MPO that carries the electric flux~\cite{Banuls:2024oxa}. The same framework yields both the forward and the vacuum-to-meson matrix elements, and our results are converged in the bond dimension and in the flux cutoff.

The two mesons display qualitatively different light-cone structure. The vector meson has a single-peaked PDF and a symmetric LCDA, while the scalar meson has a double-peaked PDF and an antisymmetric LCDA with a node at $u=1/2$ and vanishing vector-current decay constant, as required by charge conjugation. This structure is closely analogous to that of neutral scalar mesons and scalar $P$-wave quarkonia in QCD~\cite{Cheng:2005nb,Braguta:2008qe,Li:2026fgc}. For both mesons, the PDF closely
follows the normalized square of the LCDA, showing that the mesons are dominated by their valence component at the masses studied here. The remaining differences between the two distributions, most visibly the finite minimum of the scalar PDF at $\xi=1/2$, are consistent with lattice artifacts that act differently on densities and amplitudes, namely the finite correlator range and the center-of-mass motion of the meson in the open box.

Several extensions are worth further investigation. Lighter fermions, $m/g\lesssim1$, would probe the transition from quarkonium-like mesons to the strongly-coupled regime, where the scalar meson becomes a loosely bound state of two vector mesons and higher Fock components are expected to be sizable. This will require systematic continuum and infinite-volume extrapolations, as well as improved control of the residual charge-conjugation breaking of highly excited states. Beyond this, the same framework gives access to generalized parton distributions and form factors through boosted states, and to transition distributions between the vector and scalar mesons, which are allowed by charge conjugation. Finally, extending these calculations to multi-flavor and non-Abelian gauge theories in $1+1$ dimensions would give access to the light-cone structure of pion-like and baryon-like states.

\acknowledgments
We are grateful to Joao Barata, Yahui Chai, Meijian Li, Yang Li, and Manuel Schneider for fruitful discussions. This work is supported by the National Natural Science Foundation of China (NSFC) under Grant No. 12675177 (W. Q.), by new faculty startup funding from Central China Normal University (W. Q.), and by the Chinese Exchange Program (CEP) for Young International Talents under grant number JC202606020G (J. H.). This work is also supported by IGNITE 2025, funded by Maria de Maeztu Excellence Unit budget, project code CEX2023-001318-M of the Agencia Estatal de investigación (AEl, Spanish Agency for Research).

\appendix

%% ============================================================================
\section{Non-relativistic model}
\label{app:hf_model}
%% ============================================================================
In the limit of $m\gg g$, the Schwinger model can be described as nonrelativistic system where pair creation is suppressed. The mesons can be described by a single pair bound by the electric field between them~\cite{Coleman:1975pw,Coleman:1976uz}. By Gauss's law, charges $\pm g$ at separation $r$ are connected by a constant field $E=g$ with energy $\frac12E^2|r|$, which gives the linear potential $V(r)=\sigma|r|$ with string tension $\sigma=g^2/2$. The relative motion obeys
\begin{align}
    \left[-\frac{1}{m}\frac{d^2}{dr^2}+\sigma|r|\right]\psi(r)=\epsilon\,\psi(r)\,,
\end{align}
with total energy $M_h=2m+\epsilon_h$ and binding energy $\epsilon_h$. The solution to the relative two-body Schr\"{o}dinger equation are the Airy functions,
\begin{align}
    \psi_V(r)\propto\mathrm{Ai}\Big(\frac{|r|}{r_0}-a_1'\Big)\,,\qquad
    \psi_S(r)\propto\mathrm{sgn}(r)\,\mathrm{Ai}\Big(\frac{|r|}{r_0}-a_1\Big)\,,
    \label{eq:hf_psi}
\end{align}
where $r_0=[2/(mg^2)]^{1/3}$, and $a_1'\simeq 1.0188$ and $a_1\simeq 2.3381$ are the first zeros of $\mathrm{Ai}'(x)$ and $\mathrm{Ai}(x)$, respectively. The bound states alternate between even and odd wave functions, which have $\Cc=-1$ and $\Cc=+1$, respectively. Therefore, the vector meson is the ground state and the scalar meson the first excitation with
\begin{align}
    \frac{M_V}{g}=\frac{2m}{g}+a_1'\,\Big(\frac{g}{4m}\Big)^{1/3}\,,\qquad
    \frac{M_S}{g}=\frac{2m}{g}+a_1\,\Big(\frac{g}{4m}\Big)^{1/3}\,.
    \label{eq:hf_mass}
\end{align}
With $\tilde\psi_h(p)=\int dr\,e^{-ipr}\psi_h(r)$ normalized to $\int\frac{dp}{2\pi}|\tilde\psi_h|^2=1$, the PDF and LCDA are respectively
\begin{align}
    f_h(\xi)=\frac{m}{\pi}\,\big|\tilde\psi_h\big((1-2\xi)m\big)\big|^2\,,
\end{align}
and 
\begin{align}
    \phi_V(u)=\frac{m\,\tilde\psi_V\big((2u-1)m\big)}{\pi\,\psi_V(0)}\,,\qquad
    \phi_S(u)=\frac{i\,m^2\,\tilde\psi_S\big((2u-1)m\big)}{\pi\,\psi_S'(0)}\,,
    \label{eq:hf_dist}
\end{align}
with the corresponding decay constants from Eqs.~\eqref{eq:fV} and \eqref{eq:scalar_norm},
\begin{align}
    \mathsf{f}_V=\frac{\psi_V(0)}{\sqrt m}\simeq0.624\Big(\frac{g}{m}\Big)^{1/3}\,,\qquad
    \mathsf{f}_S=\frac{|\psi_S'(0)|}{m^{3/2}}=\frac{g}{2m}\,.
    \label{eq:hf_decay}
\end{align}
These equations provide a good approximation to the theory in the non-relativistic limit; however, it fails when $m\lesssim g$.

%% ============================================================================
\section{Numerical convergence}
\label{app:convergence}
%% ============================================================================

\begin{figure}[htp]
    \centering
    \includegraphics[width=0.47\linewidth]{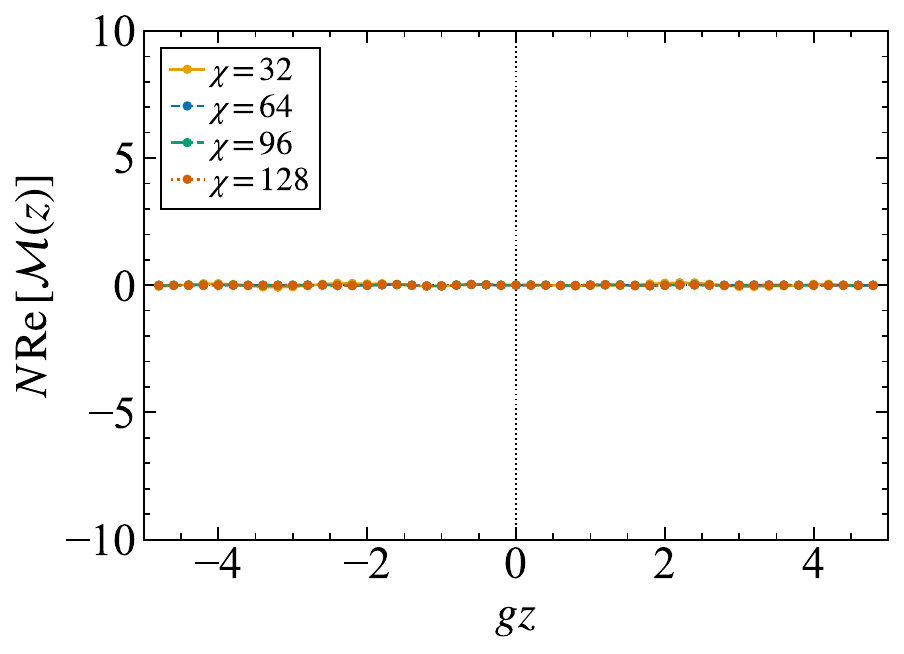}
    \hfill
    \includegraphics[width=0.47\linewidth]{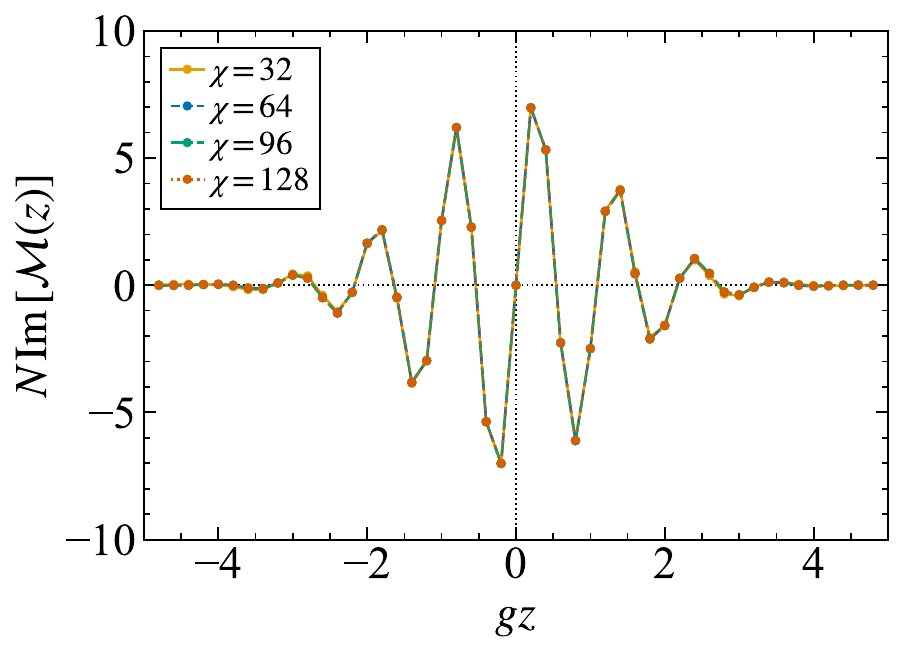}
    \includegraphics[width=0.47\linewidth]{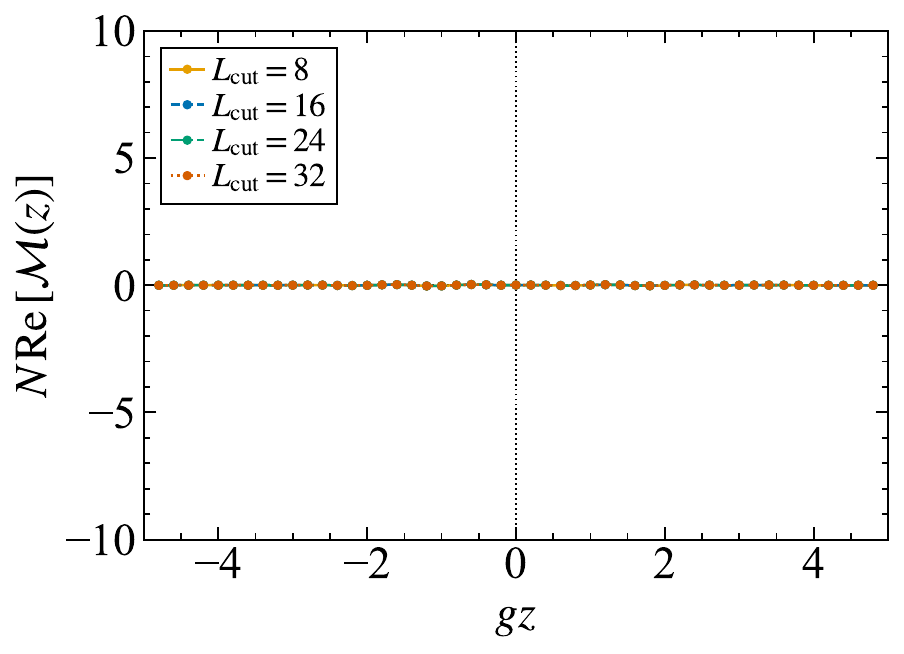}
    \hfill
    \includegraphics[width=0.47\linewidth]{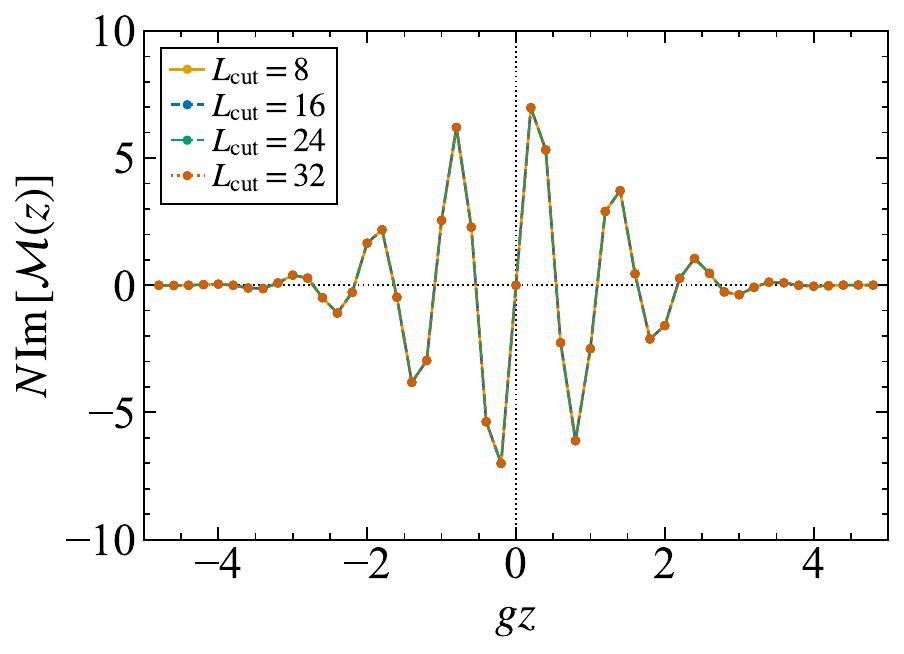}
    \includegraphics[width=0.47\linewidth]{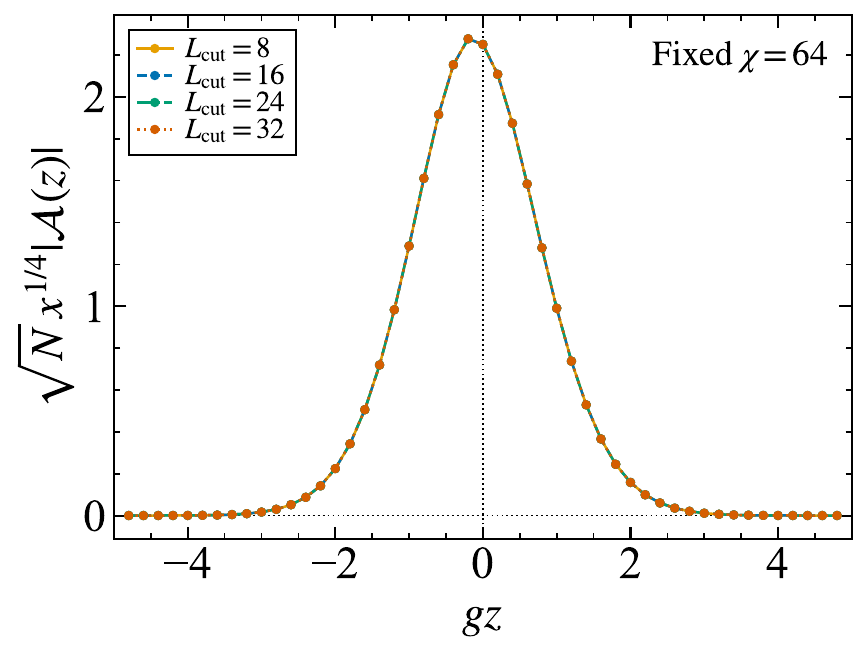}
    \hfill
    \includegraphics[width=0.47\linewidth]{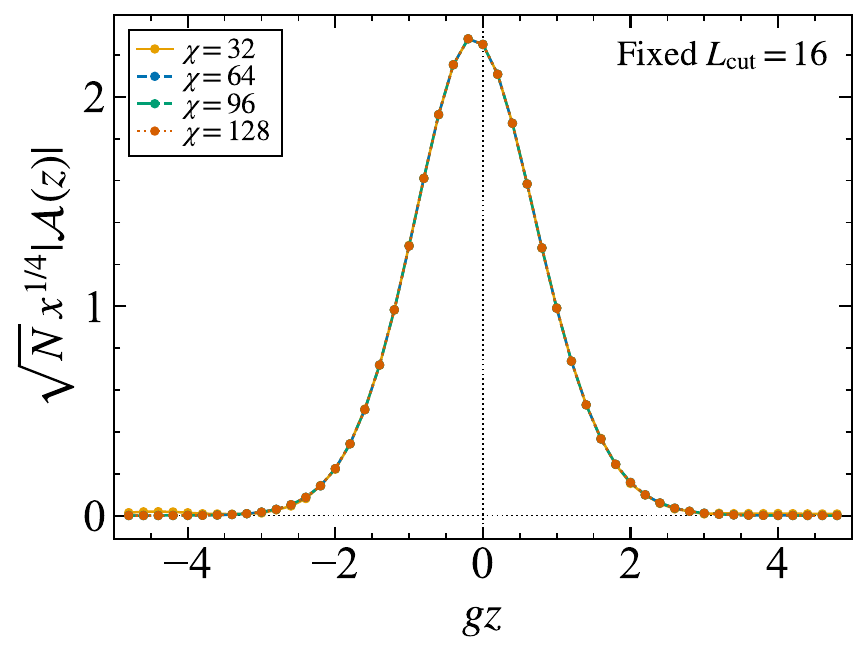}
    \caption{Convergence of the PDF and LCDA correlator of the vector mesons for $m/g=10/\sqrt{\pi}$ at fixed $\Lcut=16$ and at fixed $\bdim=64$.
    }
    \label{fig:conv_me_vector}
\end{figure}

\begin{figure}[htp]
    \centering
    \includegraphics[width=0.47\linewidth]{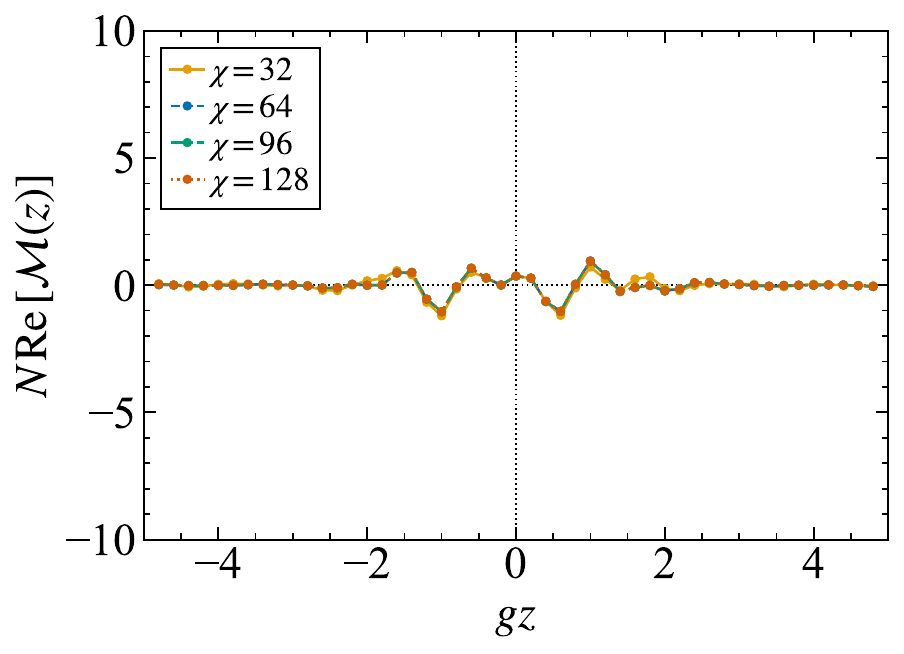}
    \hfill
    \includegraphics[width=0.47\linewidth]{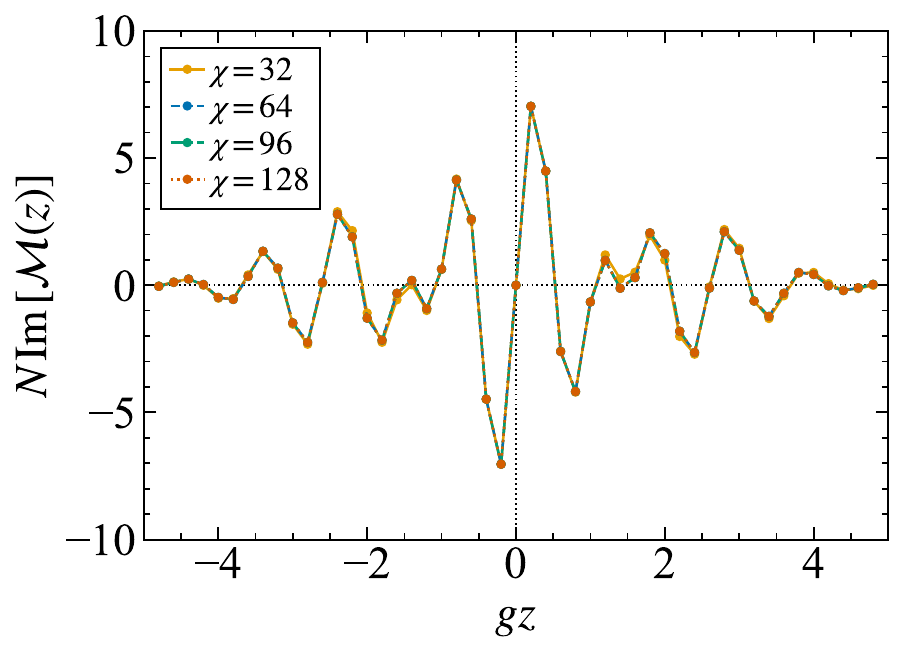}
    \includegraphics[width=0.47\linewidth]{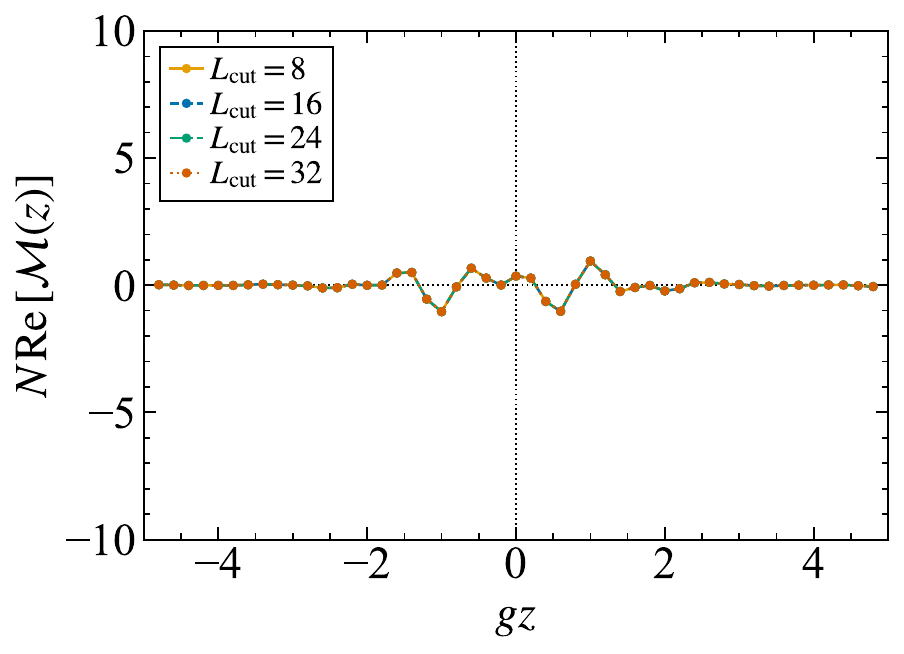}
    \hfill
    \includegraphics[width=0.47\linewidth]{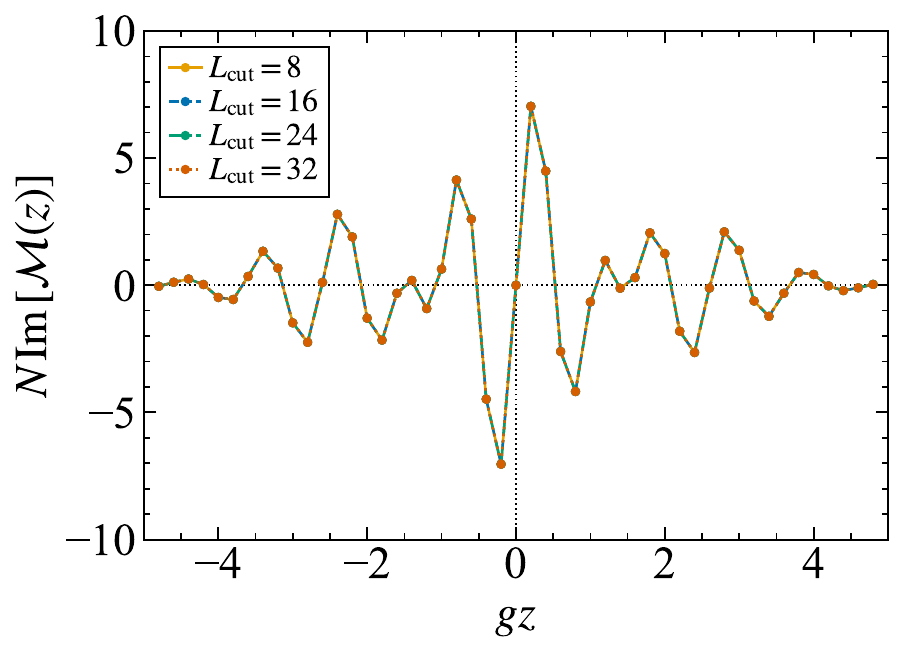}
    \includegraphics[width=0.47\linewidth]{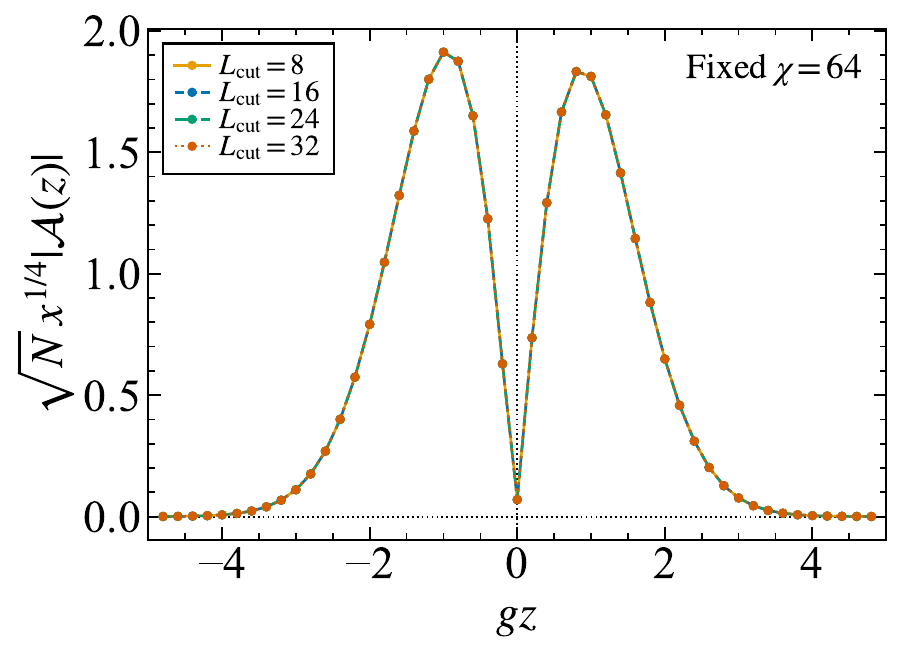}
    \hfill
    \includegraphics[width=0.47\linewidth]{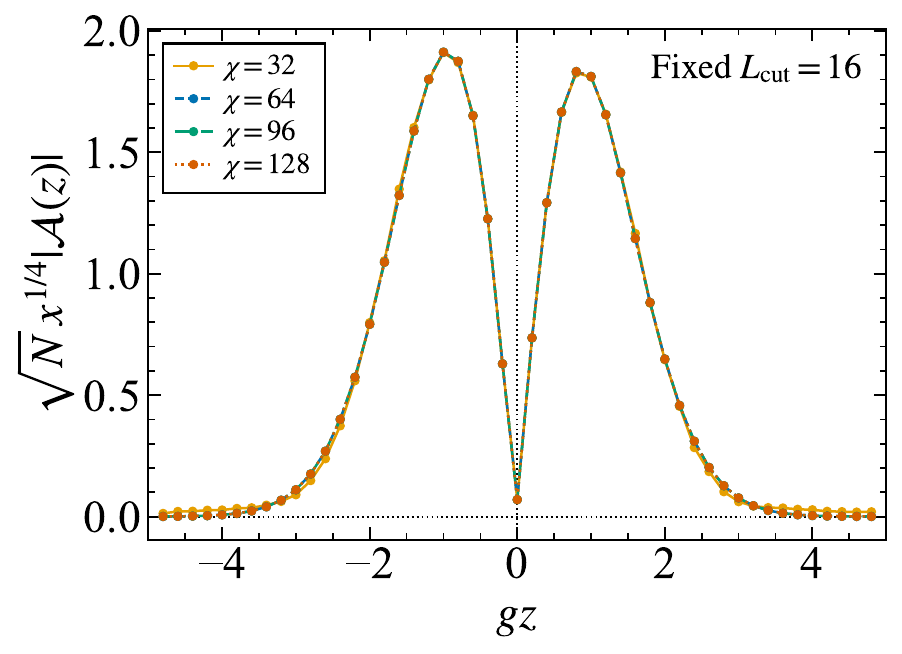}
    \caption{Convergence of the PDF and LCDA correlator of the scalar mesons for $m/g=10/\sqrt{\pi}$ at fixed $\Lcut=16$ and at fixed $\bdim=64$.
    }
    \label{fig:conv_me_scalar}
\end{figure}

Here, we verify the two truncation parameters of the time evolution: the maximum bond dimension $\bdim$ and the flux cutoff $\Lcut$. Figures~\ref{fig:conv_me_vector} and~\ref{fig:conv_me_scalar} show the PDF correlators of the vector and scalar mesons for $\bdim\in\{32,64,96,128\}$ at $\Lcut=16$ and for $\Lcut\in\{8,16,24,32\}$ at $\bdim=64$, for $m/g=10/\sqrt{\pi}$. The results are indistinguishable on the scale of the plots, so we used $\bdim=64$ and $\Lcut=16$ for the main results in the paper. The rapid convergence in $\bdim$ reflects the limited entanglement generated during the evolution time $t\le z_{d_{\max}}$, and the insensitivity to $\Lcut$ shows that large electric fluxes are not populated. 

\begin{figure}[htp]
    \centering
    \includegraphics[width=0.47\linewidth]{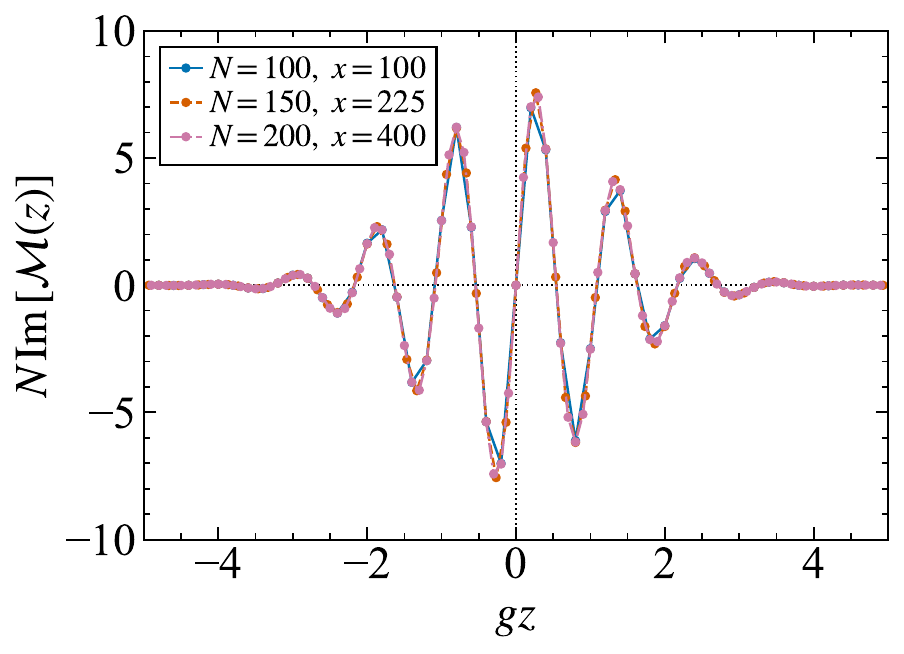}
    \hfill
    \includegraphics[width=0.47\linewidth]{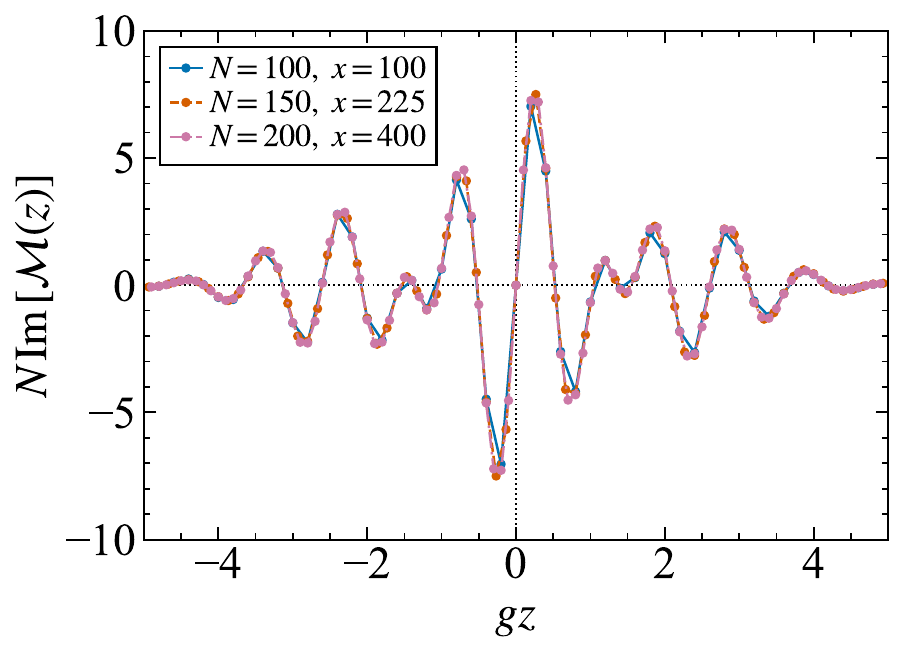}
    \includegraphics[width=0.47\linewidth]{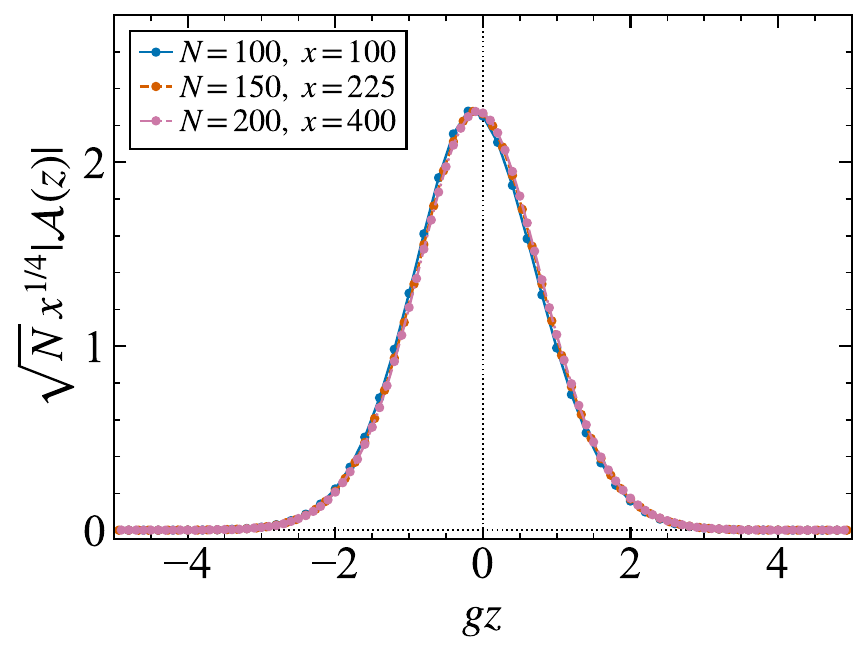}
    \hfill
    \includegraphics[width=0.47\linewidth]{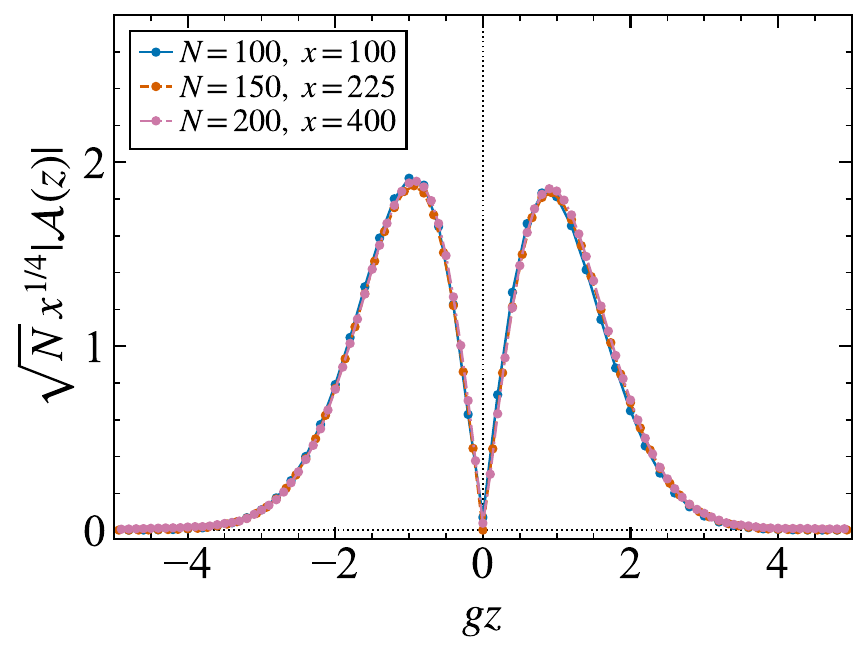}
    \includegraphics[width=0.47\linewidth]{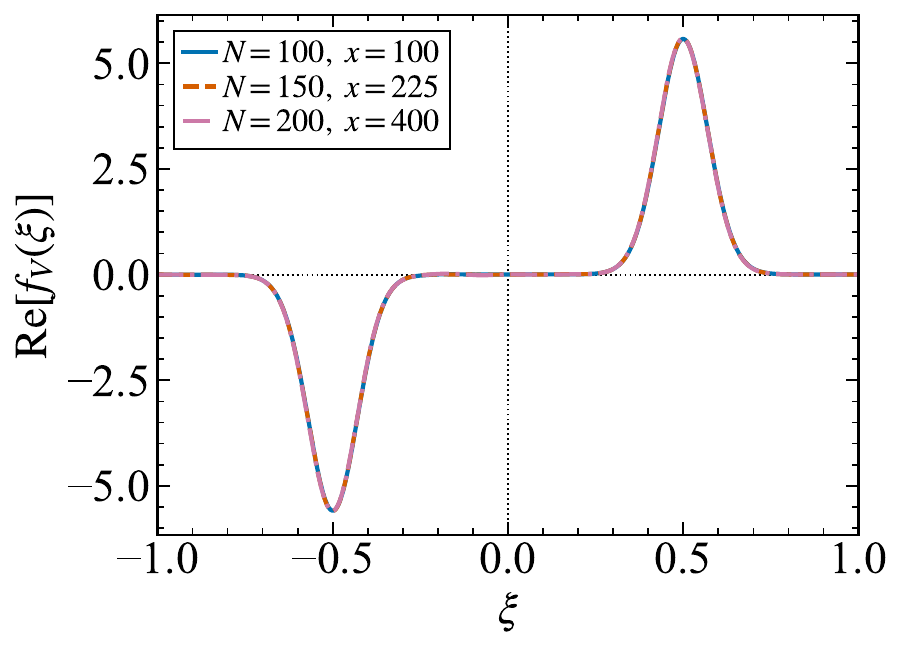}
    \hfill
    \includegraphics[width=0.47\linewidth]{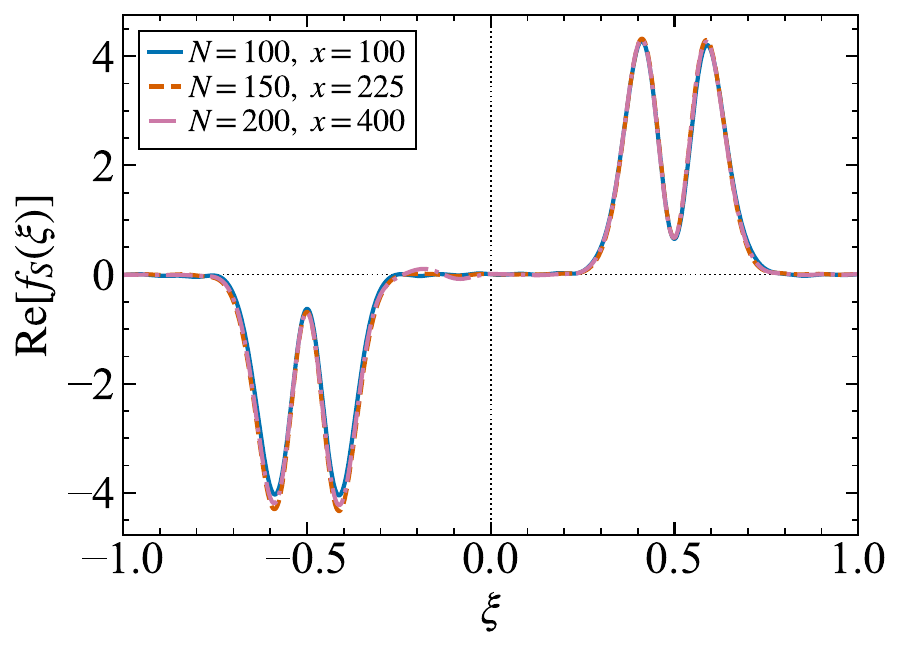}
    \includegraphics[width=0.47\linewidth]{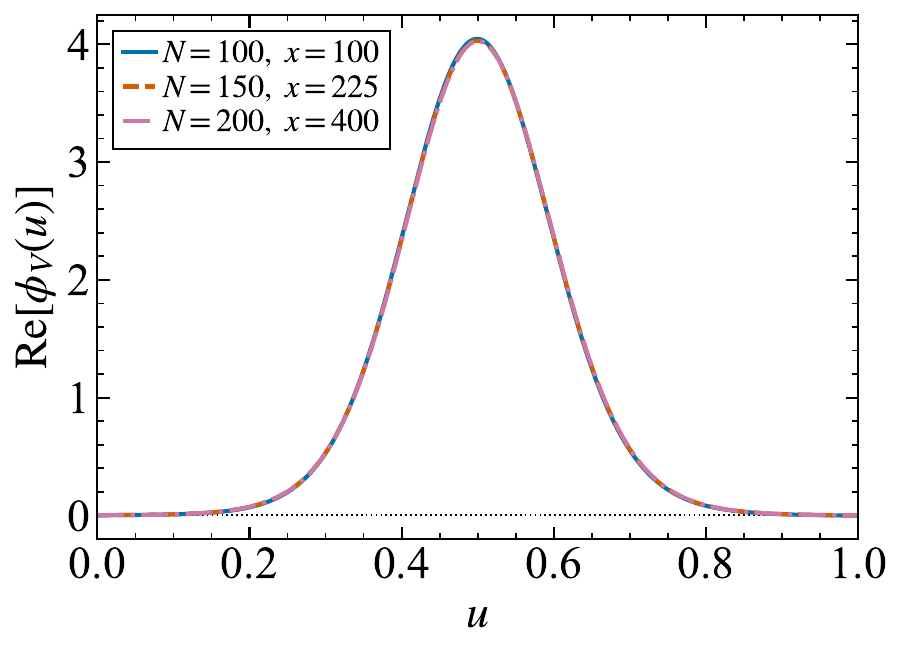}
    \hfill
    \includegraphics[width=0.47\linewidth]{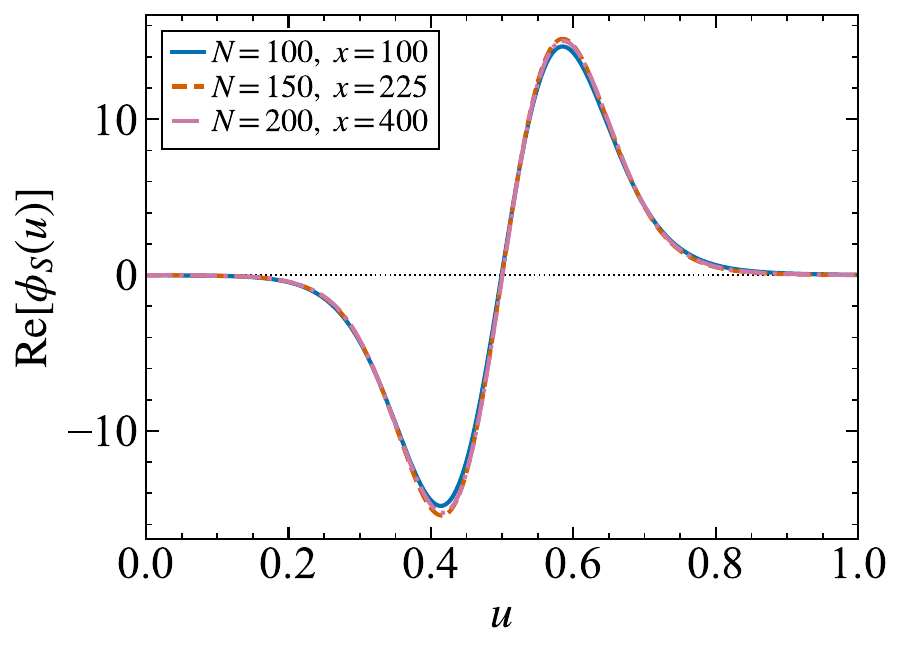}
    \caption{Convergence of the PDF and LCDA correlator and distribution for various lattice-spacing at fixed physical volume $\mathcal{V}=10$ for $m/g=10/\sqrt{\pi}$. Three sets of lattice spacings are presented: $x=100, 225, 400$ (or $ag=0.1, 0.067, 0.05$) with $N=100, 150, 200$ respectively.}
\label{fig:conv_fixed_volume}
\end{figure}

We also examine the lattice-spacing dependence at fixed physical volume $\mathcal{V}=10$. Figure~\ref{fig:conv_fixed_volume} compares the PDF correlators, LCDA correlators, PDFs, and LCDAs of both mesons for $(N,x)=(100,100)$, $(150,225)$, and $(200,400)$ at $m/g=10/\sqrt{\pi}$. Increasing $x$ reduces the lattice spacing, $ag=1/\sqrt{x}$, while increasing $N$ keeps the physical volume unchanged. The distributions show only small changes over this range, indicating weak lattice-spacing dependence for these parameters. This comparison does not assess finite-volume effects, which require futher investigation of $\mathcal{V}$ dependence.

\bibliographystyle{JHEP-2modlong.bst}
\bibliography{Lib.bib}

\end{document}